\documentstyle[12pt,epsf]{article}
\newcommand{\la}[1]{\label{#1}}
\newcommand{\be}{\begin{equation}}
\newcommand{\ee}{\end{equation}}
\newcommand{\ba}{\begin{eqnarray}}
\newcommand{\ea}{\end{eqnarray}}

\newcommand{\bi}{\begin{itemize}}
\newcommand{\ei}{\end{itemize}}
\newcommand{\rmi}[1]{{\mbox{\scriptsize #1}}}
\newcommand{\fig}{Fig.~}
\newcommand{\nr}[1]{(\ref{#1})}

\newcommand{\hphi}{{\hat{\phi}}}

\newcommand{\bfx}{\mbox{\bf x}}
\newcommand{\fr}[2]{{\frac{#1}{#2}}}
\newcommand{\msbar}{\overline{\mbox{\rm MS}}}
\newcommand{\half}{{\scriptstyle{1\over2}}}

\def\lsi{\raise0.3ex\hbox{$<$\kern-0.75em\raise-1.1ex\hbox{$\sim$}}}
\def\gsi{\raise0.3ex\hbox{$>$\kern-0.75em\raise-1.1ex\hbox{$\sim$}}}

\newcommand{\gsim}{\mathop{\gsi}}

\makeatletter \@addtoreset{equation}{section} \makeatother
\renewcommand{\theequation}{\arabic{section}.\arabic{equation}}

\begin{document}

\begin{titlepage}
\begin{flushright}
CERN-TH/97-332\\
hep-lat/9711048\\
November 1997\\
\end{flushright}
\begin{centering}
\vfill

{\bf Three-dimensional U(1) gauge+Higgs theory 
as an effective\\ theory for finite temperature
phase transitions}
\vspace{0.8cm}

K. Kajantie$^{\rm a,b}$\footnote{keijo.kajantie@cern.ch},
M. Karjalainen$^{\rm b}$\footnote{mika.karjalainen@helsinki.fi},
M. Laine$^{\rm a}$\footnote{mikko.laine@cern.ch},
J. Peisa$^{\rm c}$\footnote{peisa@amtp.liv.ac.uk} \\

\vspace{0.3cm}
{\em $^{\rm a}$Theory Division, CERN, CH-1211 Geneva 23,
Switzerland\\}
\vspace{0.3cm}
{\em $^{\rm b}$Department of Physics,
P.O.Box 9, 00014 University of Helsinki, Finland\\}
\vspace{0.3cm}
{\em $^{\rm c}$Department of Mathematical Sciences, 
University of Liverpool, \\ Liverpool L69 3BX, UK}

\vspace{0.7cm}
{\bf Abstract}

\end{centering}

\vspace{0.3cm}\noindent
We study the three-dimensional U(1)+Higgs theory (Ginzburg-Landau
model) as an effective theory for finite temperature phase transitions
from the 1 K scale of superconductivity to the relativistic scales of
scalar electrodynamics. The relations between the parameters of the
physical theory and the parameters of the 3d effective theory are
given. The 3d theory as such is studied with lattice Monte Carlo
techniques. The phase diagram, the characteristics of the transition
in the first order regime, and scalar and vector
correlation lengths are determined. We find that even rather 
deep in the first order regime, the transition is  
weaker than indicated by 2-loop perturbation theory. Topological effects 
caused by the compact formulation are studied, and it is 
demonstrated that they vanish in the continuum limit. In particular,  
the photon mass (inverse correlation length) is observed to 
be zero within statistical errors in the symmetric phase, 
thus constituting an effective order parameter. 
\vfill
\noindent

\end{titlepage}

\section{Introduction}

Finite temperature phenomena are very important for cosmology and for
the physics of ultrarelativistic heavy ion collisions. It is thus 
indispensable to have efficient and accurate methods for computing, 
e.g., the partition function for a given theory. This paper is
devoted to solving this problem for the
U(1) gauge+Higgs theory, or scalar electrodynamics, 
or the Ginzburg-Landau model.
This theory in three dimensions
is the effective theory of superconductivity~\cite{kleinert}
and of liquid crystals~\cite{lc} in certain regimes, and it is also
an interesting theoretical laboratory for studying the 
thermodynamics of and topological 
defect formation in relativistic field theories.

A finite temperature system singles out a specific Lorentz frame, the
rest frame of the matter. In non-relativistic condensed matter physics
this automatically leads to a 
three-dimensional (3d) --- or in some cases even lower dimensional --- 
formulation of thermodynamical computations. The Hamiltonian is given and the
partition function is computed in 3d. For realistic systems
the full computation is usually too complicated 
and one has to replace the original theory by an effective one
focussing on the essential degrees of freedom. For instance,  
in superconductivity the
full quantum theory of electrons in an ionic lattice is first replaced
by the BCS theory, which then for a class of phenomena can be replaced
by the very simple Ginzburg-Landau model \cite{kleinert}. 

The situation is different in finite temperature relativistic field theories.
The general first-principles Lorentz and gauge invariant formulation 
is necessarily four-dimensional (4d). However, even here, for a class of
theories and phenomena, the full 4d theory can be replaced by 
effective 3d theories of various degrees of simplicity. The original
idea~[3--5] 
is rather old and has recently
been applied to a large class of relativistic field theories,
such as QCD~[6--9], 
the SU(2)+fundamental Higgs theory
or the Standard Model~[10--13], 
the Minimal 
Supersymmetric Standard Model~[14--17], 
and the SU(5)+adjoint 24-plet Higgs theory \cite{rajantie5}. The
corresponding 3d effective theories have been numerically
studied for QCD in~[19--21,9], 
for SU(2)+fundamental Higgs in~[10,22--27],
and for SU(2)$\times$U(1) + fundamental Higgs in \cite{su2u1}. The
present case of the U(1)+Higgs theory, for which 
dimensional reduction has been considered in~[12,29--31]
has been numerically studied in~[32--38]. 

The purpose of this paper is to study the U(1)+Higgs theory in
the effective theory approach, emphasizing
the two essential but totally separate aspects of the problem:
\bi

\item 
The numerical study of the 3d effective theory as such: the
determination of the phase diagram, of the characteristics 
of the phase transitions found, and of the correlation lengths of 
various gauge invariant operators. We have presented  
our first results for the correlation length measurements and 
for the structure of the phase diagram already in~\cite{u1prb}, 
and here we concentrate on the characteristics of the phase 
transition in the 1st order regime and on 
explaining the details
of the correlation length measurements.
Earlier numerical work has appeared in~[32--37]. 
The full solution of the
problem requires this numerical work, although analytic 
methods~[39--44]
give important insights, as well. 
It is also quite interesting to compare the 
U(1)+Higgs theory with several
related theories with a massless photon in the tree-level
spectrum: SU(2)$\times$U(1) + fundamental Higgs theory \cite{su2u1},
SU(2)+adjoint Higgs theory \cite{ph,adjoint} and the pure U(1)
theory~[45--47]. 
The topological mass created for 
the photons in the compact 
formulation~[48--51] 
will play an important role in this context. 

\item 
The analytic derivation of the relation between the full physical 
theory and the effective theory. Here many different physical theories
can be mapped onto one effective theory. 
The U(1)+Higgs theory is particularly
interesting in that the full theory can be, e.g., superconductivity
at $T\sim$ 1 Kelvin or hot scalar+fermion plasma at 
relativistic temperatures. 
Of course, the latter case can also be approached directly from the
4d viewpoint \cite{bhw,hebecker}. 
\ei

The results for the first aspect ---
the numerical study of the 3d theory as such ---  
are contained in Sections 2--6 of this paper. We formulate the 
problem in continuum in Sec.~2, make some perturbative computations in Sec.~3, 
discretize the theory in Sec.~4, discuss the photon in the 
discretized theory and in some other 3d theories in Sec.~5, 
and present our main lattice results in Sec.~6. Since the 
second aspect of the problem, the derivation of the 3d theory, 
completely factorizes from the first aspect, we discuss the
derivation separately in Appendix~A for the 
case of superconductivity and in~Appendix~B
for the hot scalar+fermion plasma. 
We conclude in Sec.~7.

\section{3d U(1)+Higgs theory in continuum}

The 3d U(1)+Higgs theory  is a
locally gauge invariant continuum
field theory defined by the functional integral
\ba
Z&=&\int {\cal D}A_i{\cal D}\phi\,\exp\bigl[-S(A_i,\phi)\bigr], \la{z} \\
S&=&\int d^3x\biggl[\fr14 F_{ij}^2+
|D_i\phi|^2 
+m_3^2 \phi^*\phi + \lambda_3 \left(\phi^*\phi\right)^2\bigg], 
\label{ac}
\end{eqnarray}
where $F_{ij}=\partial_iA_j-\partial_jA_i$ 
and $D_i=\partial_i+ie_3A_i$.
The parameters $m_3,e_3^2,\lambda_3$ of the Lagrangian
have the dimension GeV and the fields have dimension GeV$^{1/2}$. 
Since the theory in eq.~\nr{ac} is a continuum field
theory, one has to carry out ultraviolet renormalization. In
3d the couplings $e_3^2$ and $\lambda_3$ are not renormalised
in the ultraviolet, but
there is a linear 1-loop and a logarithmic 2-loop 
divergence for the mass parameter $m_3^2$. 
In the $\msbar$ dimensional regularization scheme in 
$3-2\epsilon$ dimensions, the renormalized mass parameter
becomes \cite{perturbative}
\be
m_3^2(\mu) =  {-4e_3^4+8\lambda_3e_3^2-8\lambda_3^2\over16\pi^2}
\log\frac{\Lambda_m}{\mu},
\la{m3}
\ee
where $\Lambda_m$ is a scale independent physical mass parameter
of the theory. Instead of it, it is more convenient to use
$m_3^2(e_3^2)$. Choosing $e_3^2$ to set the scale,
the physics of the theory will depend on the two dimensionless
ratios
\be
y= { m_3^2(e_3^2)\over e_3^4},\quad x={\lambda_3\over e_3^2}.
\la{parameters}
\ee
We shall later in Appendices~A and~B
relate the parameters of superconductivity
and finite $T$
scalar electrodynamics to $x$ and $y$.

\begin{figure}[t]
\hspace{1cm}
\epsfxsize=8cm
\centerline{\epsffile{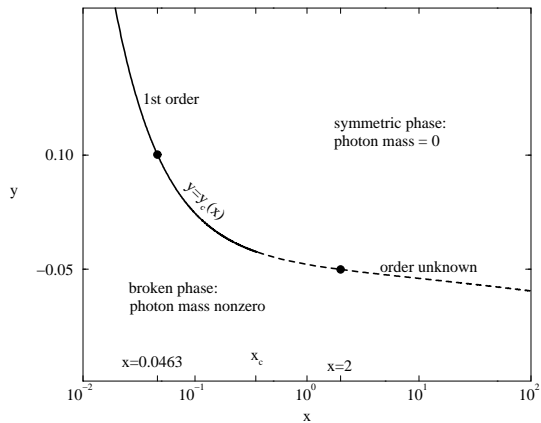}}

\vspace*{-0.5cm}

\caption[a]{The phase diagram of the G-L theory. The photon 
mass acts as an effective order parameter.}
\la{ycx}
\end{figure}

It is now a mathematical and computational problem to determine the
physics associated with the theory in eq.~\nr{z}. As usual all physics lies
in expectation values of various operators. Since this is a gauge
theory, only gauge invariant operators have non-vanishing expectation
values. The most relevant of these are the local (depending only
on one point $\bfx$) operators of lowest
dimensionalities (half-odd dimensions could be made integer
by multiplying by $e_3$):
\bi

\item Dim = 1: the $J^{PC}=0^{++}$ scalar 
$S(\bfx)=\phi^*(\bfx)\phi(\bfx)$.

\item Dim = 1$\fr12$: the $1^{+-}$ vector 
$\tilde V_i(\bfx)\equiv B_i=\fr12 \epsilon_{ijk}F_{jk}(\bfx)$.

\item Dim = 2: the $1^{--}$ vector 
$V_i(\bfx)={\rm Im}\,\phi^*(\bfx)D_i\phi(\bfx)$;  the $1^{-+}$ vector
$S_i(\bfx)={\rm Re}\,\phi^*(\bfx)D_i\phi(\bfx)=\partial_i
\phi^*(\bfx)\phi(\bfx)$;
and the $0^{++}$ scalar 
$(\phi^*\phi)^2$.

\item Dim = 2$\fr12$: the $1^{+-}$ vector 
$\phi^* B_i\phi$.

\item Dim = 3: The $0^{++}$ scalars $F_{ij}F_{ij}$ and
$\phi^* D_iD_i\phi$;  and the $2^{++}$ tensor
$\phi^*[ \{D_i,D_j\}-2/d\,\, \delta_{ij}D_kD_k]\phi$; etc.

\item Dim = 3$\fr12$: The $0^{--}$ scalar $B_i\,\partial_i
\phi^*\phi$; and the $0^{-+}$ scalar $B_i\,
{\rm Im}\,\phi^* D_i\phi$; etc.

\item Dim = 4: The $0^{+-}$ scalar $\partial_i\phi^*\phi\,
{\rm Im}\,\phi^* D_i\phi$; etc.

\ei
The quantum numbers here refer to O(3). From these one can further
construct, e.g., bilocal operators such as correlators of two
of the above operators acting at different points.

The first topic of importance is the phase structure of the theory.
The critical curve $y=y_c(x)$ (see Fig.~\ref{ycx}) divides
the plane into two disjoint regions, the symmetric phase at
$y>y_c(x)$ and the broken phase at $y<y_c(x)$. 
The presence of a critical curve
is signalled by singularities in the free energy
\be
Z=\exp\bigl[-Ve_3^6 f(y,x)\bigr],
\ee
where $f$ is dimensionless. The critical curve 
is localised by measuring 
different local or bilocal
expectation values of the gauge invariant
local operators listed above.
There 
is no local 
gauge invariant order parameter, the expectation value of which
would vanish in one of the phases. Instead, we shall use as an
effective order parameter the photon mass, which is measured from
a bilocal operator.

For small $x$, the system has a first-order
transition (Fig.~\ref{ycx}). It then has two bulk states of
the same free energy, and the broken and symmetric
phases coexist at $y_c$. 
The system has one stable and one metastable branch for 
$y_-(x)<y<y_+(x)$.
Expectation values of various gauge invariant scalar operators
jump when crossing $y_c(x)$. The following
quantities are of particular interest:
\begin{itemize}
\item The jump $\Delta\ell_3$ of the order parameter like quantity
$\ell_3\equiv\langle\phi^\dagger\phi(e_3^2)\rangle/e_3^2$ between
the broken and symmetric phases at $y_c$. In perturbation
theory,
$\Delta\ell_3\sim\half\phi^2_b(y_c)/e_3^2$, where $\phi_b$ is
the location of the broken minimum in, say, the Landau gauge.
Note that $\langle\phi^\dagger\phi(\mu)\rangle$ is scale dependent
\cite{framework}, as given below in eq.~(\ref{condensate}).
\item The interface tension $\sigma_3\equiv\sigma/(T e_3^4)$, 
defined in perturbation theory by
\be
\sigma_3=\int_0^{\phi_b/e_3}\,d(\phi/e_3)
\sqrt{2V(\phi/e_3)/e_3^6},
\la{sigmapert}
\ee
where $V$ is the perturbatively computed 3d effective potential. On
the lattice $\sigma_3$ is determined from the fundamental relation
$F=-pV+\sigma A$, which holds in the presence of a planar interface
separating two coexisting phases.
\ei
In a second order transition, on the other hand,
the quantities of interest are the
different critical exponents. 

The second essential property of the U(1)+Higgs theory 
is its excitation spectrum. In perturbation theory, one can easily 
see what the free field degrees of freedom are both in the 
symmetric and broken phases (see also Table 1). 
However, the symmetric phase is in some respects
non-perturbative (since the Coulomb potential is logarithmically
confining in 2+1 dimensions) and thus 
these states need not be the physical ones. The physical 
states are described by the correlators of the gauge invariant operators
listed above. For example, the physical mass,  
or inverse correlation length, of the lowest scalar 0$^{++}$ excitation
is obtained from the large distance behaviour of the correlator
$\langle S(\bfx)S({\bf y})\rangle$. To study vector 1$^{--}$ excitations
we shall use the operators $\tilde V_i(\bfx), V_i(\bfx)$.  
In condensed matter context these
masses are usually called ``renormalised" masses; we prefer to 
reserve the word renormalisation for the elimination of ultraviolet 
divergences.

A particularly interesting state is the photon. It is
evidently one of the fundamental fields of the action in the 
symmetric phase; is it in the physical spectrum of the theory?
In~\cite{u1prb} we have
provided non-perturbative evidence for the fact that it 
indeed is. It can thus serve as an effective order parameter and 
the phase diagram can contain two disconnected regions as in Fig.~1.
This is a nontrivial fact, since massless states can disappear from
the physical spectrum or in the lattice formulation of the theory.
This will be discussed in more detail in Section 5.

\section{Perturbative results for the U(1)+Higgs theory}

Let us consider the region of small $x$. Then 
perturbation theory is expected to be accurate and it 
predicts a ``fluctuation induced" or ``Coleman-Weinberg-type" first order
phase transition. Renormalised perturbation theory induces a dependence
on a scale $\mu$ and in 3d one has to go to 2 loops in order to 
optimize this scale. The 2-loop potential has been discussed in
\cite{hebecker,perturbative,jpp,kp} and optimisation
in \cite{perturbative,kp}, and we refer
there for details (see also eq.~(\ref{V2sum}) below for the
2-loop potential in the full 4d theory with fermions). 

To show the main qualitative features
of the perturbative results, 
consider the 1-loop potential in
the Landau gauge (see eq.~\nr{1looppot}),
\be
V_\rmi{1-loop}/e_3^6 =
\biggl\{\fr12y(\mu)\hphi^2+\fr14x\hphi^4-{1\over12\pi}
\Bigl[2\hphi^3+(y+3x\hphi^2)^{3/2}+(y+x\hphi^2)^{3/2}-2y^{3/2}\Bigr]
\biggr\},
\la{v1loop}
\ee
where $\hphi=\phi/e_3$ and
\be
y(\mu)=y+{1\over16\pi^2}(-4+8x-8x^2)\ln{e_3^2\over\mu}.
\ee
The $\mu$ dependence of $y(\mu)$ 
is of 2-loop order so that it is a higher order effect
in eq.~\nr{v1loop}.
The leading $x\to0$ result is obtained by including only
the 1-loop vector term, the first 1-loop term
in eq.~(\ref{v1loop}). The potential is then, equivalently,
\be
V_{\rm 1-loop}/e_3^6\approx
\fr14x\hphi^2\biggl[\biggl(\hphi-{1\over3\pi x}
\biggr)^2+{2y\over x}\biggl(1-{1\over18\pi^2xy}\biggr)\biggr].
\la{vonlyvector}
\ee
Two degenerate states are obtained when the last term vanishes.
{}From this one finds that
\ba
y_c(x)&=&{1\over18\pi^2x},\quad y_+(x)={1\over16\pi^2x},
\quad y_-(x)=0,\la{yc2}\\
\hphi_\rmi{symm}&=&0,\quad\hphi_\rmi{broken}={1\over3\pi x},\\
\Delta\ell_3&=&{1\over18\pi^2x^2},\quad\sigma_3={2^{3/2}\over648\pi^3x^{5/2}}.
\la{leading}
\ea

Taking into account the other terms in eq.~\nr{v1loop} and
the 2-loop effects, one can improve on the accuracy.
Perturbative results for the critical curve, the latent heat and
the interface tension in different approximations 
are shown in
Figs.~\ref{xy} and \ref{Dl3}. The curves are scaled with the leading
$x\to0$ dependences (eqs.~\nr{yc2}, \nr{leading}). The variation as
a function of $x$ in the 1-loop results is due to the scalar
loop terms in eq.~\nr{v1loop}. The 2-loop results have been computed
by either optimising \cite{perturbative,kp} the scale $\mu$ or by
fixing it to be $\mu= e_3^2$. The scale dependence gives an idea
of the magnitude of higher order corrections. One observes that
perturbation theory 
is becoming unreliable 
at least for $x\gsim0.1\ldots0.2$.

Concerning the mass spectrum, 
using the approximations $m_W=e_3\phi_b$, $m_H^2=V''(\phi_b)$, 
one has in the broken phase, 
\ba
m_W/e_3^2 &=&\left(1+\sqrt{1-16\pi^2yx}\right)/(4\pi x),\\
m_H^2/e_3^4&=&\left(1-16\pi^2yx+\sqrt{1-16\pi^2yx}\right)/(8\pi^2x).
\ea
More accurate expressions for these correlator masses as poles
of the 1-loop propagators are given in \cite{kp}. 

The mass spectrum in the symmetric phase is a more subtle issue.
The photon is argued to be massless to all orders
in perturbation theory~\cite{hebecker,blaizot}. Nevertheless, 
non-perturbative effects can produce a mass in the discretized
theory, see Sec.~5. The scalar excitation, on the other hand,
is represented by a bound state in a logarithmically confining
potential and thus the computation of its mass is non-trivial.
A simple way to see this is to consider the
3d theory in Minkowskian terms, so that 
the excitation masses are masses 
of states living in two space dimensions
(note that the masses are not those of 
bound states in 3d). 
The scalar excitation then corresponds to 
a $\phi^+\phi^-$ bound state in a 2d Coulombic potential
\be
V_\rmi{Coul}(r)={e_3\over2\pi}K_0(\mu r)\sim{e_3\over2\pi}\log
{1\over\mu r} ;\la{coulpot}
\ee
$\mu\sim e_3^2$ is an infrared (IR) cutoff. 
A simple minimisation argument then shows
that the masses are of the form
\be
M = 2m_3-{e_3^2\over4\pi}\log{m_3\over e_3^2}+{\rm const}
\times e_3^2 + {\cal O}\Bigl(\frac{e_3^4}{m_3}\Bigr),
\ee
where the logarithm comes from the Coulombic term and
the constant depends on the IR cut-off $\mu$ 
and on the quantum numbers of the state. Since the constant 
is sensitive to the IR cut-off, it is to 
be determined by numerical means.

\begin{figure}[tb]

\vspace{-0.5cm}

\hspace{1cm}
\epsfxsize=9cm
\centerline{\epsffile{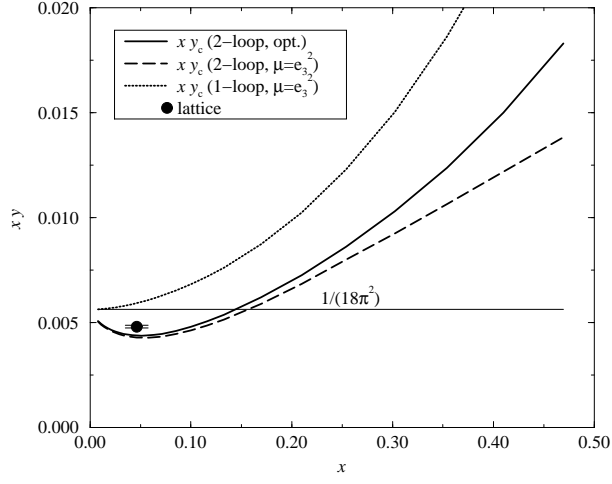}}

\vspace{-1cm}

\caption[a]{\protect The quantity $x\times y_c(x)$ computed
perturbatively for the U(1)+Higgs theory to different accuracies.
The horizontal line is the leading $x\to0$ result in eq.~\nr{yc2}.
The lattice Monte Carlo point for $x=0.0463$ is also shown. For
$x=2$ perturbation theory is meaningless; then the lattice 
number is $x\times y_c(x=2)\approx -0.10$.}
\la{xy}
\end{figure}

\begin{figure}[tb]

\vspace*{-0.5cm}

\centerline{\hspace{-3.3mm}
\epsfxsize=9cm\epsfbox{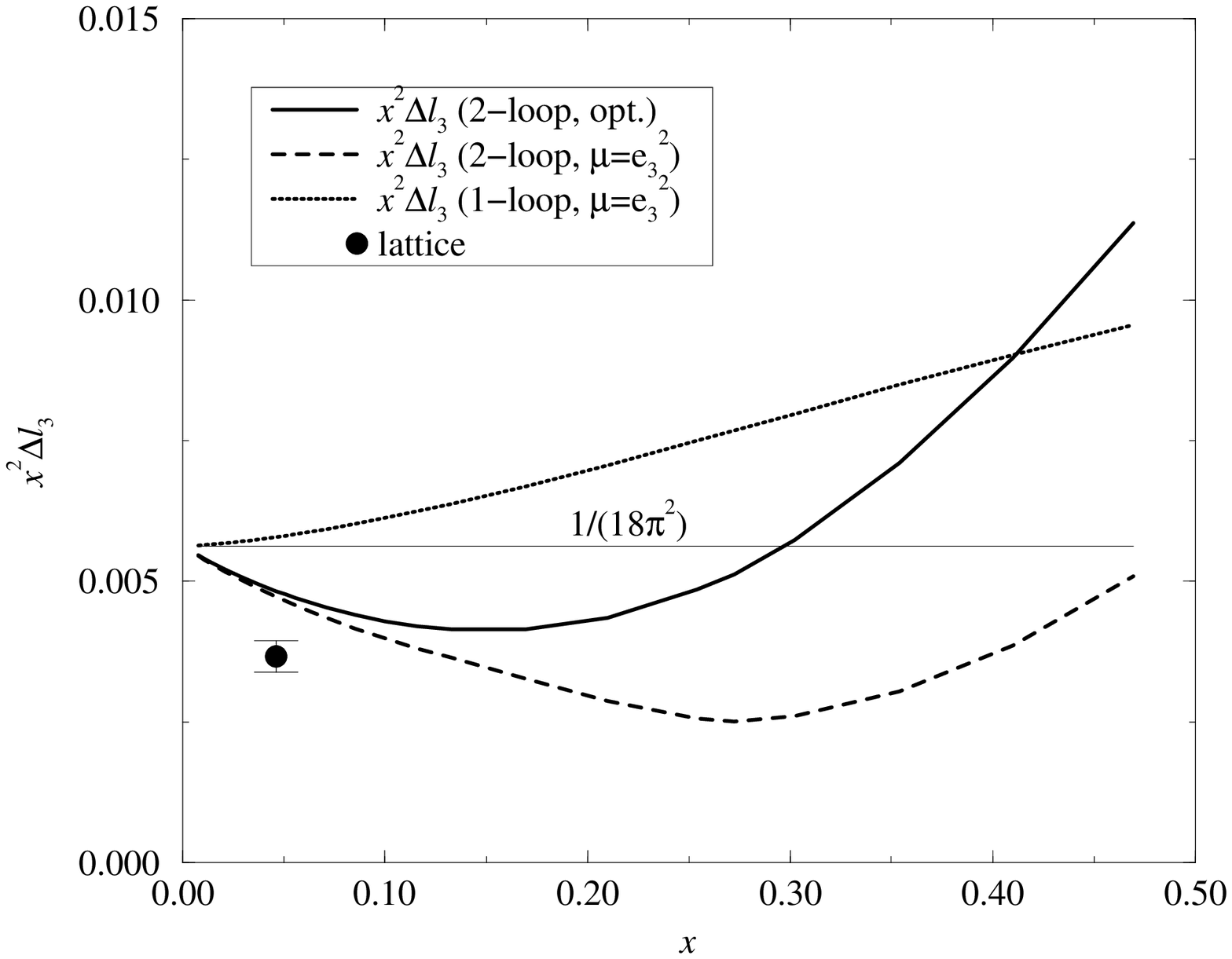}
\epsfxsize=9cm\epsfbox{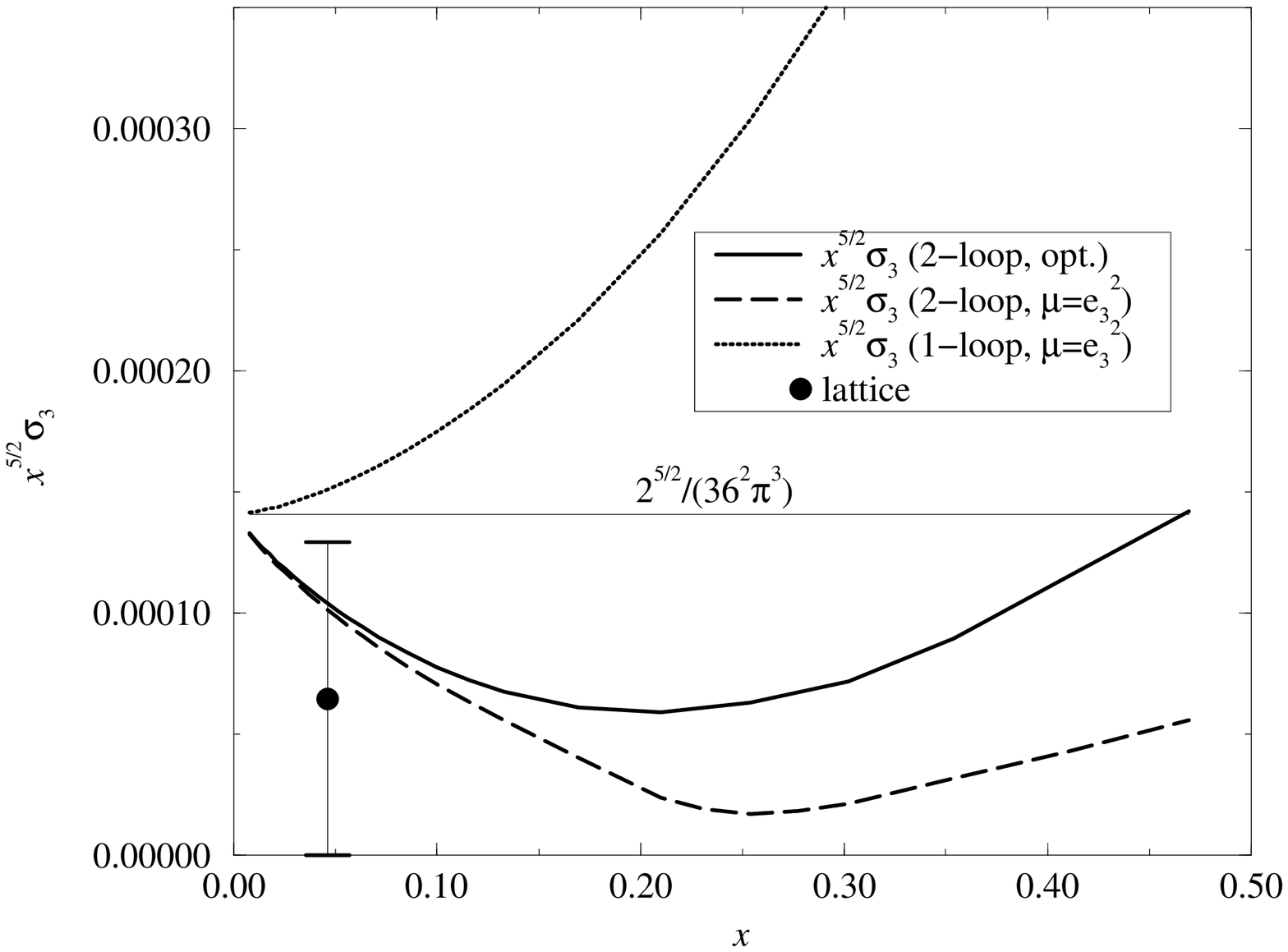}}

\vspace*{-1cm}

\caption[a]{The same as \fig\ref{xy} but for the jump of the order
parameter ($\times x^2$) and for the interface tension ($\times x^{5/2}$).}
\la{Dl3}
\end{figure}

\section{U(1)+Higgs theory on the lattice}

Our aim is to study the continuum theory in eq.~\nr{ac} and it is thus
crucial to keep the continuum variables $y,x$ fixed when
discretizing the theory. This point has to do with ultraviolet
renormalisation. The linear and logarithmic mass divergences
are renormalised in the continuum theory in the $\msbar$ scheme.
Discretization in itself is another scheme. The two schemes now
have to be related so that correlators measured in the lattice scheme 
go to the correct continuum ones at given $y,x$ in the limit
that the lattice spacing $a$ goes to zero. This computation has been carried
out in~[54--56]; 
for order ${\cal O}(a)$ corrections,
see \cite{moore_a}.

To discretize, we use the compact formulation of the gauge field
and introduce the link field $U_i(\bfx)=
\exp[iae_3A_i(\bfx)]\equiv\exp[i\alpha_i(\bfx)]$. 
Rescaling the continuum scalar field to a 
dimensionless lattice field by 
\be
\phi^*\phi\to \beta_H\phi^*\phi/2a,
\la{scaling}
\ee
the lattice action becomes 
\ba
S & = & \beta_G \!\!\sum_{\bfx,i<j}
\Bigl[1-\cos \hat{F}_{ij}(\bfx)\Bigr] 
  -\beta_H \sum_{\bfx, i} {\mbox{Re}}\, 
\phi^*(\bfx) U_i(\bfx)\phi(\bfx+\hat{i})\nonumber\\   
  & & +\sum_{\bfx} \phi^*(\bfx) \phi(\bfx)
+\beta_R\sum_{\bfx} \left[\phi^*(\bfx)\phi(\bfx)-1\right]^2, 
 \la{standardaction} 
\ea
where $\beta_G=1/e_3^2a$,
\be
\hat{F}_{ij}(\bfx)=\alpha_i(\bfx)+\alpha_j(\bfx+\hat i)-
\alpha_i(\bfx+\hat j)-\alpha_j(\bfx), \la{Fijlat}
\ee
and a specific, customary choice has been made for $\beta_H$ (other 
choices are possible as well; see~\cite{u1prb} for the action 
before any choice has been made). With these conventions, 
the lattice couplings $\beta_G, \beta_H, \beta_R$ are 
determined from~\cite{laine,lr} 
\ba
& & \beta_G={1\over e_3^2a},\qquad \beta_R={x\beta_H^2\over4\beta_G},\\
& & 2\beta_G^2\left({1\over\beta_H}-3-{x\beta_H\over2\beta_G}\right)
=y-{3.1759115(1+2x)\beta_G\over2\pi}\la{betah}\\
&&\hspace*{2cm} -{1\over16\pi^2}
\left[(-4+8x-8x^2)(\log6\beta_G+0.09)+25.5+4.6x\right].\nonumber
\ea
Thus for a given continuum theory depending on one scale $e_3^2$
and the two dimensionless 
parameters $y,x$,  the use of a lattice introduces a regulator
scale $a$, and eqs.~\nr{standardaction}--\nr{betah}  
specify, up to terms of order $e_3^2a$~\cite{moore_a}, 
the corresponding lattice action.

In a non-compact formulation of the gauge field one would replace
$1-\cos\hat F_{ij} \to \hat F_{ij}^2/2$
in eq.~\nr{standardaction}. 
Then the coefficient 25.5 in 
eq.~\nr{betah} should be replaced with -1.1~\cite{lr}.

An important quantity we will study is the scalar condensate 
$\langle\phi^*\phi\rangle$. In perturbation theory 
this has a linear and a
logarithmic divergence and renormalisation induces a known
logarithmic scale dependence $\langle\phi^*\phi(\mu)\rangle$
\cite{framework}. The condensate
itself is not a physical quantity, only its changes are. The
relation between the continuum (at the scale $\mu=e_3^2$)
and lattice condensates is \cite{lr}
\be
{\langle\phi^*\phi(e_3^2)\rangle_\rmi{cont}
\over e_3^2}=\fr12\beta_H\beta_G
\langle\phi^*\phi\rangle_\rmi{latt}-{3.1759115\beta_G\over4\pi}-
{1\over8\pi^2}\biggl[\log(6\beta_G)+0.668\biggr],
\la{condensate}
\ee
where ``latt" refers to the normalisation of the field in the
lattice action \nr{standardaction}. The first term here is the
scaling in eq.~\nr{scaling}, the second is the linear,
and the third the logarithmic divergence. The discontinuity of 
$\langle\phi^*\phi(\mu)\rangle$ in a first order transition 
is free from these divergences, 
as they are just constants.

For mass measurements we use 
the discretized forms of the operators discussed in Sec.~2.
The $0^{++}$ and $1^{-+}$ charge conjugation even ``scalar'' operators 
$\phi^*\phi$ and Re$\,\phi^*D_i\phi$ and the $1^{--}$ 
vector operators Im$\,\phi^*D_i\phi$ and $\epsilon_{ijk}F_{jk}$
are on the lattice represented by
\ba
S(\bfx)&=&\phi^{\ast}(\bfx)\phi(\bfx), \la{oo} \\
S_i(\bfx)&=& \mathop{\rm Re}\phi^{\ast}(\bfx)U_i(\bfx)\phi(\bfx+\hat i), 
\\
V_i(\bfx)&=& \mathop{\rm Im}\phi^{\ast}(\bfx)U_i(\bfx)\phi(\bfx+\hat i),
\\ 
\tilde V_i(\bfx)&=&\fr12\epsilon_{ijk}\sin\hat F_{jk}(\bfx).\la{oi}
\ea
In the 4d case these operators have been used for mass measurements
in \cite{ejjln,khmm}. Note that other (higher dimensional) operators with 
the same quantum numbers could be 
considered as well, and a  systematic way of finding 
out which combinations couple to the lowest physical mass
states, is with the blocking and consequent diagonalization techniques
discussed in Sec.~6.5. In practice, the masses are measured from 
plane-plane correlators instead of the local operators in 
eqs.~\nr{oo}--\nr{oi} as discussed in~\cite{u1prb}, and the 
photon mass measurement requires an external momentum~\cite{bpana}.

\section{Photon in 3d theories}

Various 3d continuum theories with a photon in the physical spectrum
are listed in Table 1. Here we shall add to this list the U(1)+Higgs
theory in the compact lattice formulation, eq.~\nr{standardaction}
and, for comparison, also the discretized pure U(1) gauge theory
in the compact formulation. The situation with
photons in these theories is as follows:

\begin{table}[ht]
\center
\begin{tabular}{|l|l|l|l|l|} 
\hline 
Theory & d.o.f.'s in  & d.o.f.'s in  & $\gamma$ in   & $\gamma$ in   \\
       & symm.\ ph.\ & broken ph.\ & symm.\ ph.? & broken ph.? \\
\hline
U(1)+Higgs &$A_i,\phi^*,\phi$ &$W_i,H$ & yes & no \\
           & = 2+2                &= 3+1 &       &     \\ \hline
SU(2)+adj. Higgs&$A_i^a,\phi_a$     &$W_i^\pm,A_i,H$ & no 
&yes$\to$no \\
\cite{ph,adjoint}  & = 3$\cdot$2+3  & = 2$\cdot$3+2+1 &   &   \\ \hline
SU(2)$\times$U(1)+fund. &$A_i^a,B_i,\phi_k$ &$W_i^\pm,Z_i^0,A_i,H$
& yes   &yes \\
Higgs \cite{su2u1} & = 3$\cdot$2+2+4  & = 3$\cdot$3+2+1 &      &    \\
\hline
\end{tabular}
\caption[]{3d continuum theories with a massless photon. The 2nd and
3rd columns list and count the free field degrees of freedom in the action
(with and without a shift to the broken minimum). The last columns
express whether a massless photon really appears in the physical
spectrum. For SU(2)+adjoint Higgs the perturbatively massless 
photon becomes 
massive through non-perturbative effects~\cite{polyakovphoton}. }
\end{table}

$\bullet$ 
In 3d SU(2)$\times$U(1)+fundamental Higgs theory there is
a massless field, ``photon", in both the symmetric and broken
phases. It is appears both perturbatively and non-perturbatively
as a physical state \cite{su2u1}. 
In the symmetric phase the photon is
represented by the hypercharge field, whereas
in the broken phase the photon field is a linear combination
of the hypercharge
and SU(2) gauge fields. Because the photon is massless in both phases, the
two phases cannot be distinguished by its value and the phase diagram
can be smoothly connected.

$\bullet$ 
In 3d SU(2)+adjoint Higgs theory there is, perturbatively, 
a massless photon in the broken phase but not in the
symmetric phase. This is since after symmetry breaking by
$\phi^a=(0,0,v)$ there remains a compact U(1) invariance, related to 
rotations between the 1,2 components. On the tree level the photon
could thus be used to distinguish the two phases and the phase
diagram seems to be split into two disjoint regions. However, what
happens non-perturbatively~[48--51] 
is that the
photon becomes massive due to monopole configurations. The physics
of how a gas of monopoles induces a mass is familiar from 
that of Debye screening in a plasma: 
if there is a gas of electrons ($e,T,m_e,n_e$) in
a neutralising background, there is a plasma frequency
$\omega_p^2=e^2n_e/(m_e\epsilon_0)$ and a screening length
proportional to the inverse of $\omega_p$, so that the static
photon correlation length has become
finite due to charge screening. Since the monopole charge is
$\sim\hbar/g_3$, 
where $g_3$ is the non-abelian gauge coupling, 
one similarly obtains
\be
m_\gamma^2\sim {1\over g_3^2}n_\rmi{monopole}.
\la{mgamma}
\ee
One can further carry out a semiclassical computation \cite{ole}
leading to an expression of the type (the exponent is here for $x=0$)
\be
n_\rmi{monopole}\sim\exp\biggl[-{4\pi m_W\over g_3^2}\biggr].
\la{semiclass}
\ee
Here $m_W$ is the perturbative mass of the the states $W^\pm$ (Table 1),
and one should keep in mind that they are
actually not physical states (since they are charged).
Anyway, the fundamental quantity is the photon mass 
in eq.~\nr{mgamma}
which can be non-zero and 
has to be measured numerically. This has been done
in~\cite{ph,adjoint} and the phase diagram has been shown 
to be smoothly connected.

$\bullet$ In 3d U(1)+Higgs theory or in pure U(1) gauge theory
on the lattice in the compact formulation
there also is a massive photon. A semiclassical 
computation for this Polyakov mass gives in the weak coupling
limit $\beta_G\gg1$ \cite{bmk,aho}
\be
{m_\gamma^P}/{e_3^2}
=\pi(2 \beta_G)^{3/2}\exp\biggl[-{3.176\pi\over4}\beta_G\biggr]. 
\la{pmass}
\ee
In the strong coupling limit of small $\beta_G$, on the other hand~\cite{aho},
\be
{m_\gamma^P}/{e_3^2}\approx {\pi^2\over2}\beta_G^2\log{2\over\beta_G}.
\la{pmassstrong}
\ee
Hence even in the compact U(1)+Higgs case, 
the phase diagram is connected for all
finite $e_3^2a=1/\beta_G$, and the phase transition only appears in the
limit $\beta_G\to\infty$. Below we measure $m_\gamma$ explicitly
with a series of finite $\beta_G$'s and verify this behaviour.
Thus, the massless state which appears in perturbation 
theory in the symmetric phase remains there in the 
full non-perturbative case in the continuum limit, as well, 
dividing the phase diagram into two disjoint regions.

\section{Simulations}

In this Section we discuss the 
numerical results of our lattice simulations. 
In Sec.~6.1 we discuss the parameter values used, 
in Sec.~6.2 the location of the phase transition point, 
in Sec.~6.3 the latent heat of the transition, 
in Sec.~6.4 our estimates for the interface tension, 
in Sec.~6.5 mass measurements and in Sec.~6.6 critical exponents.

\subsection{Parameter values}
We want to focus our attention on two separate regions: that of small
$x$ (eq.~\nr{parameters}), where perturbation theory 
is expected to work and the transition is of
first order, and that of large $x$~\cite{u1prb}, 
where perturbation theory breaks down.
In terms of the original physical theories these would correspond
to type I superconductors or small Higgs masses and to type II 
superconductors or large Higgs masses, respectively.
Some guidance concerning the limiting value is obtained from the
fact that for the other 3d theories studied~\cite{isthere?,ph} 
the first order transition
terminates at $x=x_c\sim 0.1...0.3$ and that $x=0.5$ is the 
tree-level limiting value between type I and II superconductors.

Since the study of this problem is rather demanding in computer
resources, we have chosen for the simulations two values of $x$, 
$x=0.0463$ and $x=2$, corresponding to strongly type I and type II
superconductors. When mapped to hot scalar electrodynamics, these
correspond to $m_H=30, 160$ GeV.

One of the most essential points of the present 
lattice simulations is that the aim is to obtain results for the 
continuum theory \nr{ac} at fixed $y,x$. The extrapolation 
to the continuum limit thus has to be carried out carefully:
one first takes a fixed $a$ (fixed $\beta_G$, eq.~\nr{betah}) and
extrapolates to infinite volume $V\to\infty$, and then one 
extrapolates $a\to 0$. 
To estimate the required lattice sizes of $V=(N a)^3$ and of the  
lattice spacing $a$, one notes that the lattices must be
big enough to contain all relevant correlation lengths, and fine enough
so that all the relevant correlation lengths are longer than 
several lattice spacings. Consider a system 
with a typical correlation length $\xi$. Then, accordingly, 
one has to satisfy 
(on a periodic lattice) $a\ll\xi\ll Na/2$ or, in physical units,
\be
e_3^2a = 1/\beta_G\ll e_3^2\xi \ll N/(2\beta_G).
\ee
Thus we have a lower limit for $N$:
\be
N\gg e_3^2\xi\cdot 2\beta_G,
\la{nlimit}
\ee
and a lower limit for $\beta_G$:
\be
\beta_G\gg 1/e_3^2\xi.
\la{bglimit}
\ee
Eq.~\nr{nlimit} expresses quantitatively why it is difficult to
approach the continuum limit $\beta_G\to\infty$: the lattice size $N$ must
be increased simultaneously.
We shall use $\beta_G\ge4$ and find that typically
$e_3^2\xi $ varies between 0.3 and 2. Our lattice sizes
are thus chosen as
$N=32,\ldots,64$.
The lattices used are shown in Table~\ref{table:lattices}

\begin{table}[ht]
\newcommand{\tube}[2]{~~$#1^2\times #2$}
\newcommand{\cube}[1]{~~$#1^3$}
\newcommand{\mc}{_m}
\center
\begin{tabular}{|c|c|lll|} \hline
 $x$ & $\beta_G$ &  \multicolumn{3}{c|}{volumes}  \\
\hline
 0.0463  &  4 &  \tube{8}{40} & \tube{12}{60} &  \\
         &    & \tube{16}{80\mc} & &\\
         \cline{2-5}
         &  6 &  \tube{8}{40} & \tube{12}{60} &  \\
         &    & \tube{16}{80\mc} & \tube{24}{90\mc} & \tube{32}{90\mc} \\
         &    &  \tube{64}{128} &  &  \\
         \cline{2-5}
         &  8 & \tube{12}{60} & \tube{16}{80} & \\
         &    &  \tube{24}{90\mc} & \tube{32}{90\mc} & \tube{40}{120\mc} \\
         \cline{2-5}
         & 12 &  \tube{16}{80} & \tube{24}{90} & \tube{32}{90}   \\
        \hline
 2       &  4 &  \cube{32} & \cube{48} & \tube{60}{120} \\
         \hline
\end{tabular}
\caption[0]{
The lattice sizes used for the simulations at the transition
temperature for each ($x,\beta_G$)-pair. In all cases,
several $\beta_H$-values were used around the transition point. 
Multicanonical simulations are marked with the subscript
($\mc$).}\la{table:lattices}
\end{table}

For each lattice listed in Table~\ref{table:lattices} we have
performed simulations with several values of $\beta_H$. Typically at
$x=0.0463$ each lattice has 3-5 different values of $\beta_H$ and at
$x=2$ somewhat more. For $x=0.0463$ where we expect to find a first
order transition, the different values of $\beta_H$ are joined
together with the Ferrenberg-Swendsen multihistogram
method~\cite{fs}. This was not done for the $x=2$ data. The total number
of combined heat-bath/overrelaxation sweeps is between $50 000$ and
$150 000$ for each $\beta_H$ value. The measurements were not performed
after every sweep in all cases, so the statistical sample sizes range
between $20 000$ and $100 000$.

In the first order regime one expects to find supercritical slowing down
when one goes to large enough lattices. We find that for
lattice sizes larger than $V/e_3^6 \approx 300$ one has to use
the multicanonical
algorithm to ensure that a correct statistical sample is generated.

The simulations were carried out with a Cray C94 at the Finnish 
Center for Scientific Computing. The total amount of computer
power used was about 2500 h of CPU time or about
$4\times10^{15} $ floating point operations $\approx$ 130 Mflops 
year. 

\subsection{The critical point}

\subsubsection{Type I, $x=0.0463$}

In principle, if the lattice spacing is small enough, the
critical point can be obtained directly from Monte Carlo data by
observing where the mass of the lowest vector excitation $m_\gamma$
vanishes. However, in practice the mass measurements
are rather demanding in the vicinity of the critical point, and if
possible we prefer to work with local operators. There
are no known gauge invariant
local order parameters, so we use order parameter -like
quantities, which display a discontinuity at the transition point. The
actual operators used are $R^2 \equiv \phi^*\phi$ and the hopping term 
\be
L\equiv{1\over 3}\sum_{i=1}^3 \mbox{Re }\phi^*(x)U_i(x)\phi(x+\hat{i}),
\ee
averaged over the volume.

For each individual lattice and $(\beta_G, x)$ pair we locate the
pseudocritical point $\beta_{H,c}$ using several different methods: 

1. ``Equal weight'' of the histogram $p$ of $R^2$.

2. ``Equal weight'' of the histogram $p$ of $L$.

3. Maximum of the susceptibility $\chi$ of $R^2$ .

4. Maximum of the susceptibility $\chi$ of $L$.

5. Minimum of the 4th order Binder cumulant $B$ of $L$.

\noindent
We give the susceptibility in dimensionless units as
\be
\chi_{R^2} =  e_3^2 V \left\langle (\overline{\phi^*\phi}
     - \langle \overline{\phi^*\phi}\rangle)^2\right\rangle=
N^3{\beta_H^2\over4\beta_G}\left\langle (\overline{R^2}
     - \langle \overline{R^2}\rangle)^2\right\rangle, \la{susc}
\ee
where the overbar denotes a volume average, 
$\overline{R^2} = N^{-3}\sum_x R^2(\bfx)$. 
The Binder cumulant $B$ has the standard
definition.


\begin{figure}

\vspace*{-1cm}

\centerline{\hspace{-2mm}
\epsfxsize=6.5cm\epsfbox{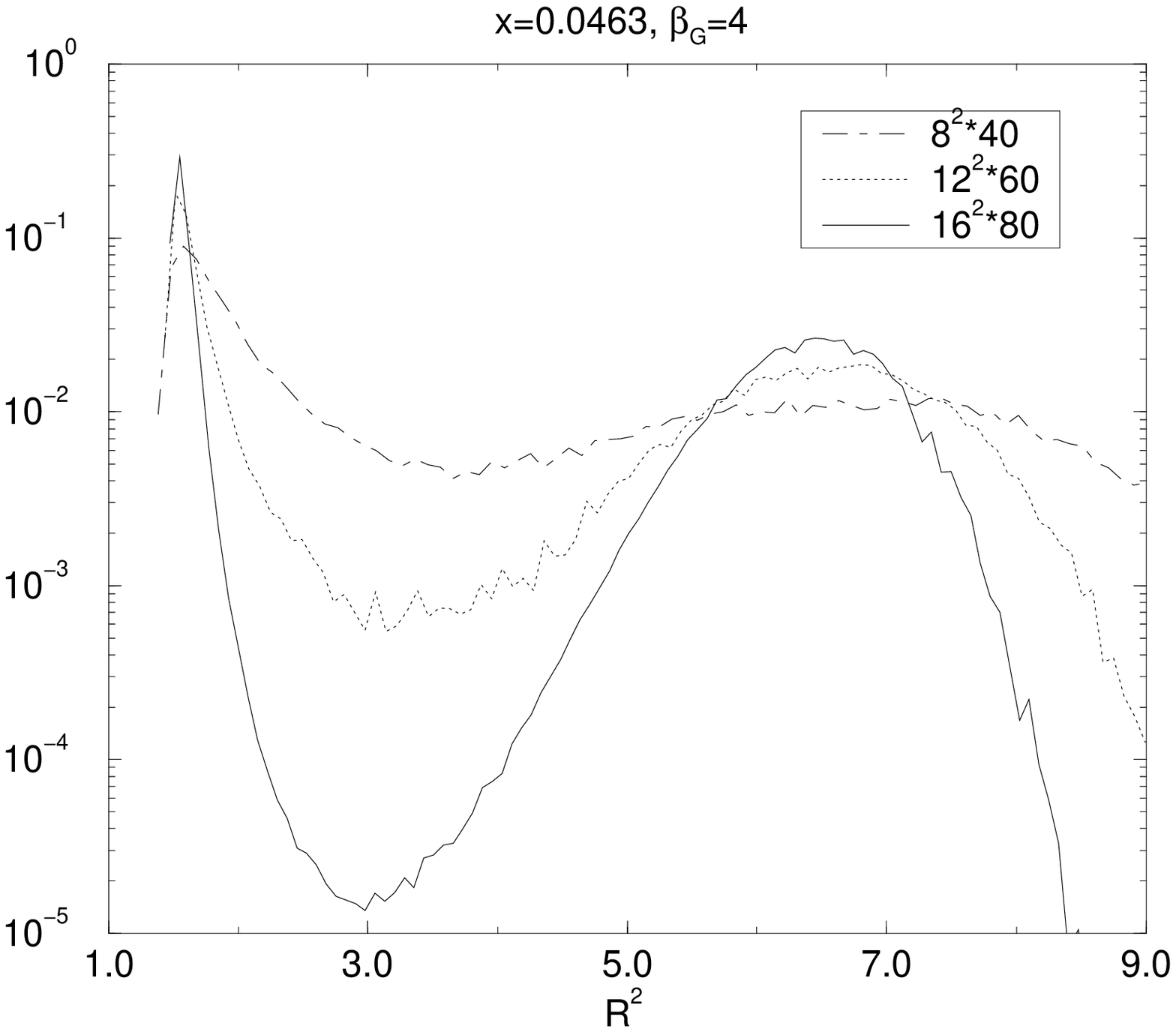}
\hspace{0cm}
\epsfxsize=6.5cm\epsfbox{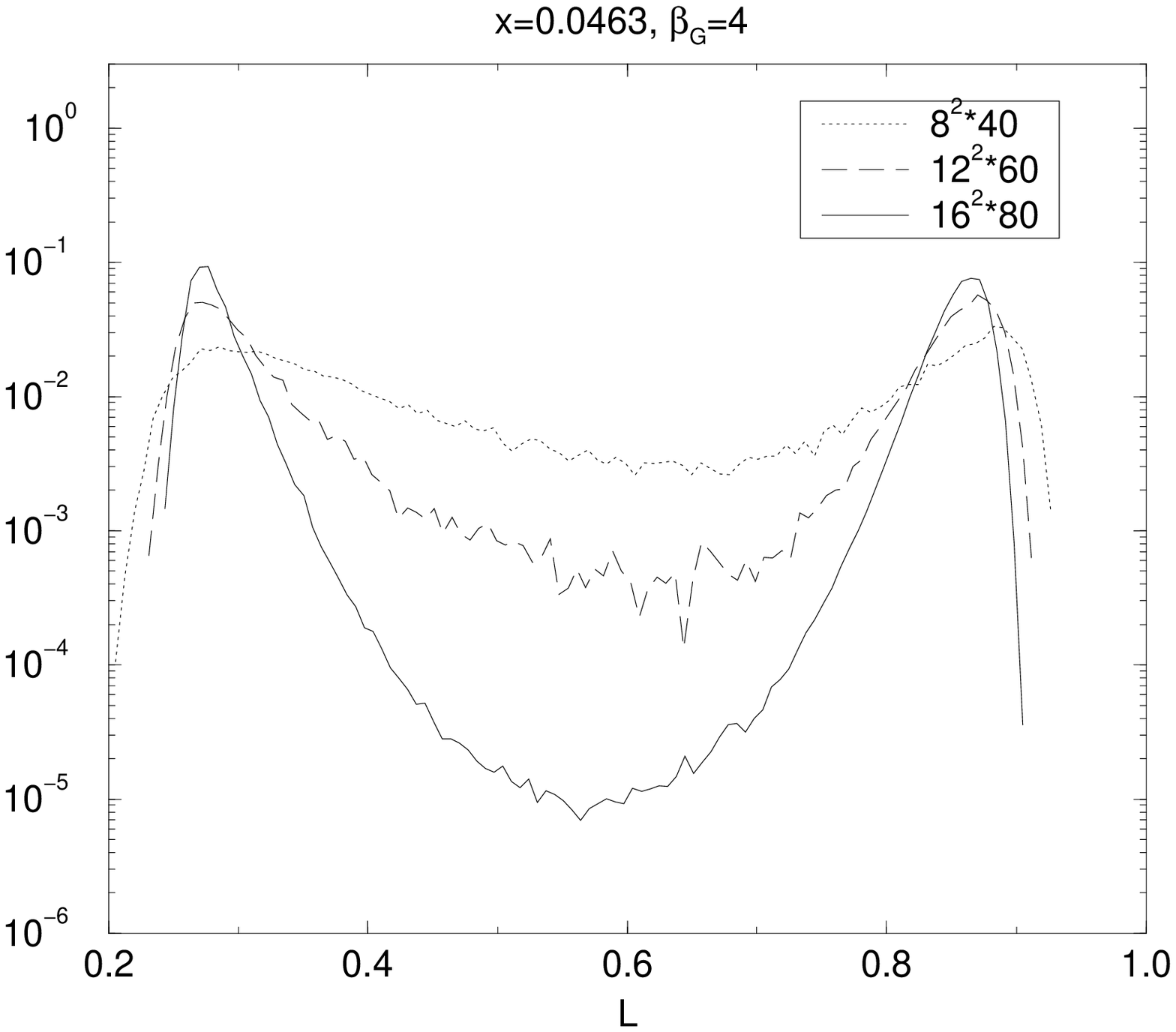}}
 
\vspace*{-0.3cm}

\centerline{\hspace{-2mm}
\epsfxsize=6.5cm\epsfbox{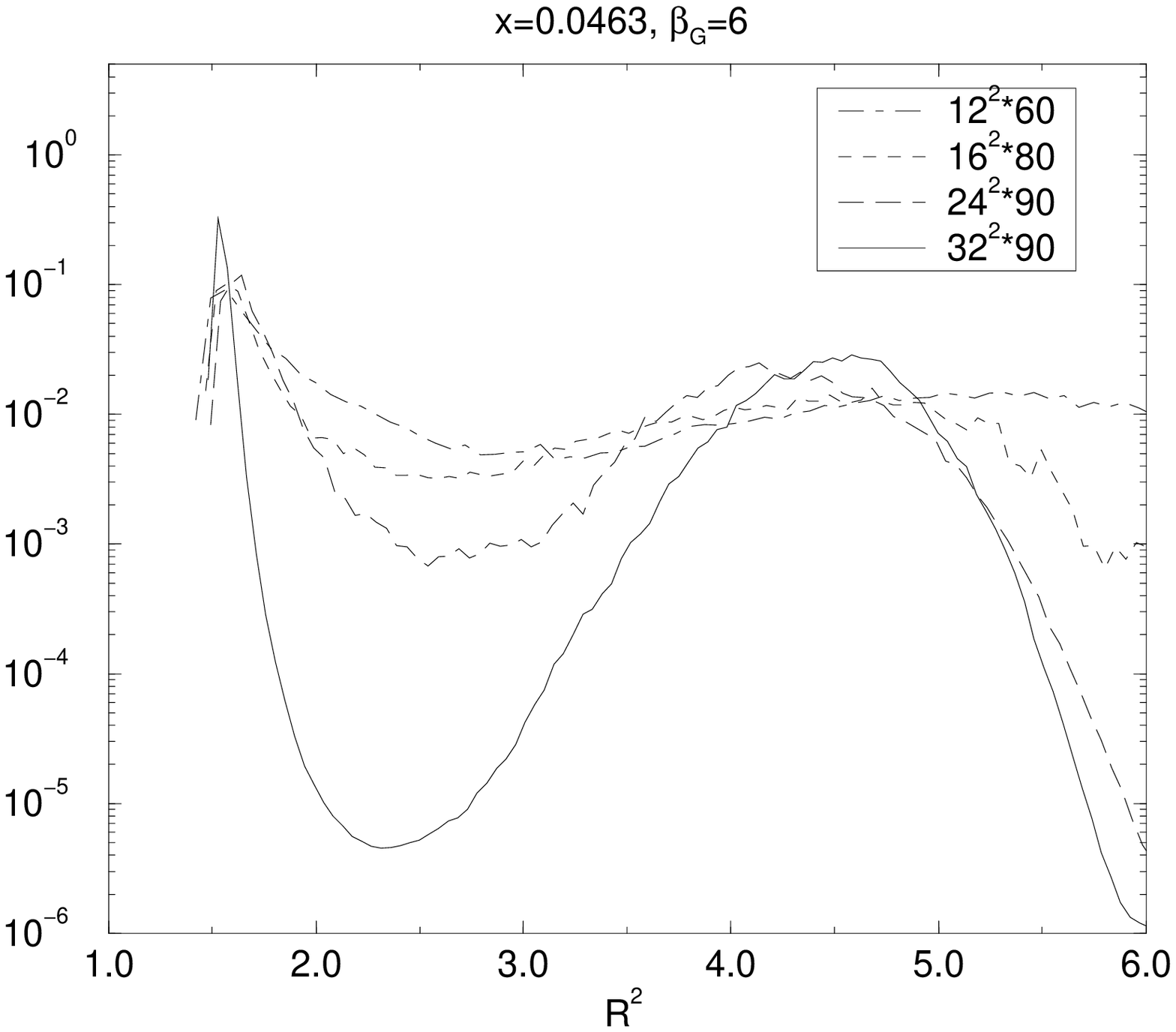}
\hspace{0cm}
\epsfxsize=6.5cm\epsfbox{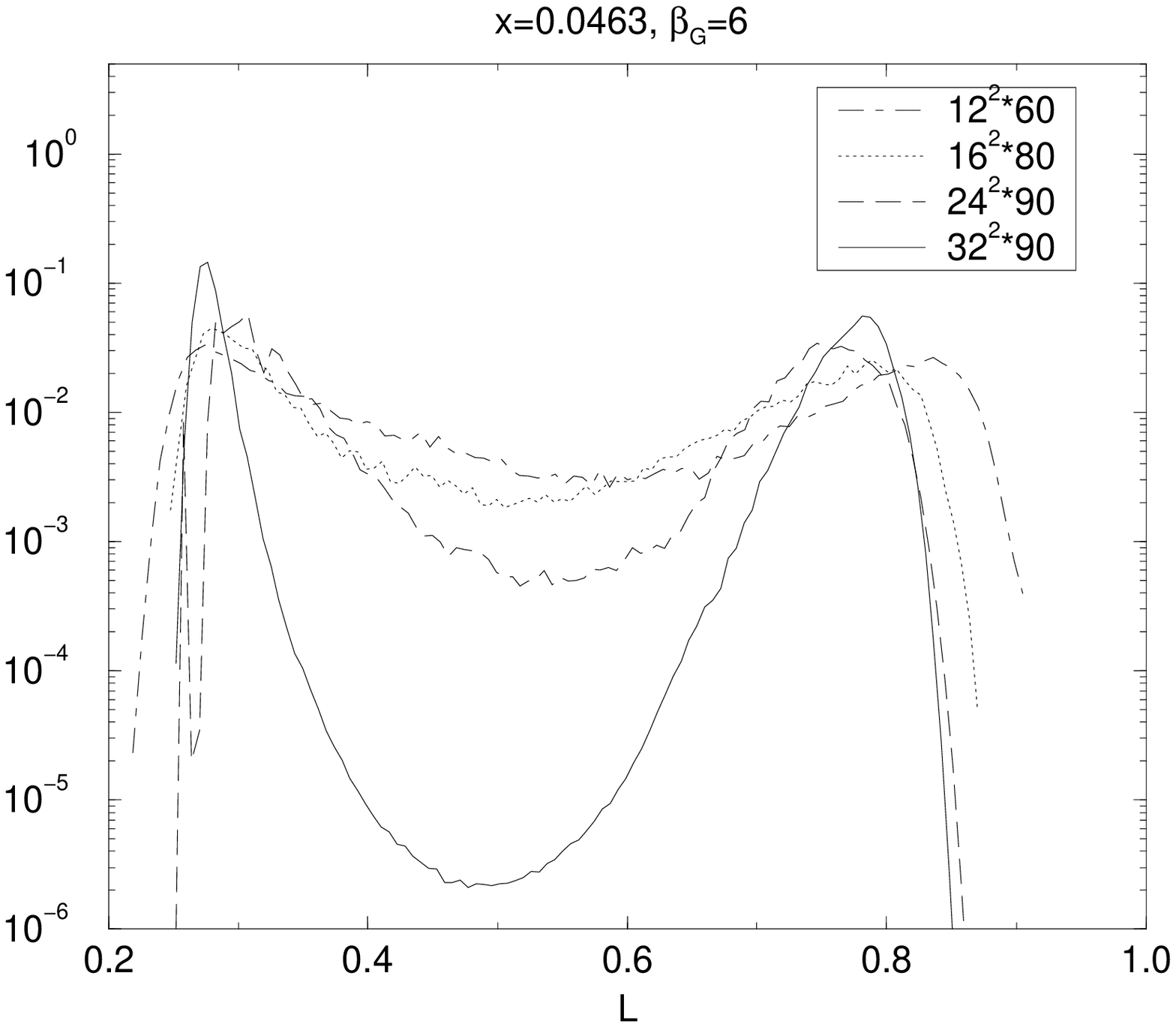}}
 
\vspace*{-0.3cm}

\centerline{\hspace{-2mm}
\epsfxsize=6.5cm\epsfbox{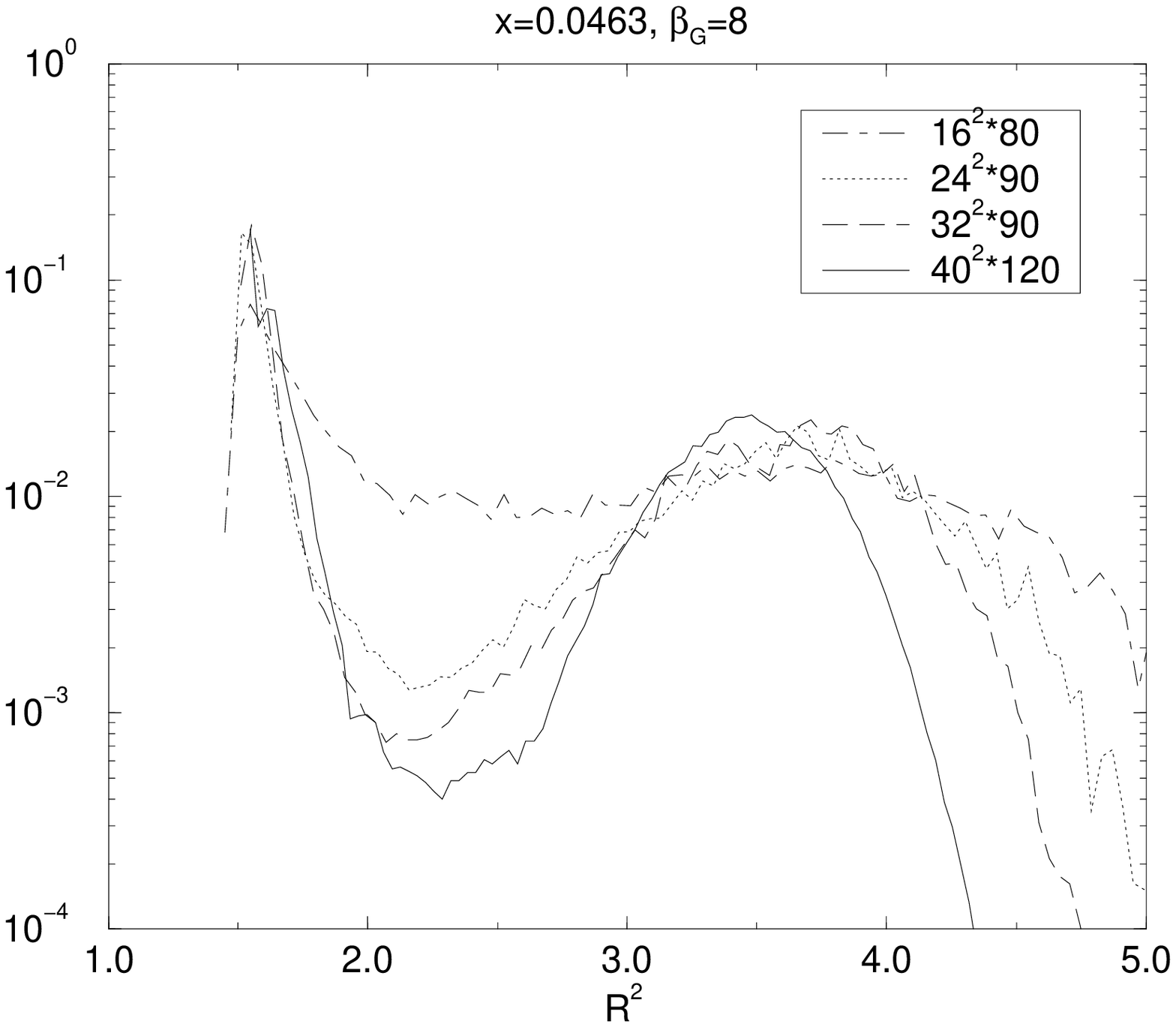}
\hspace{0cm}
\epsfxsize=6.5cm\epsfbox{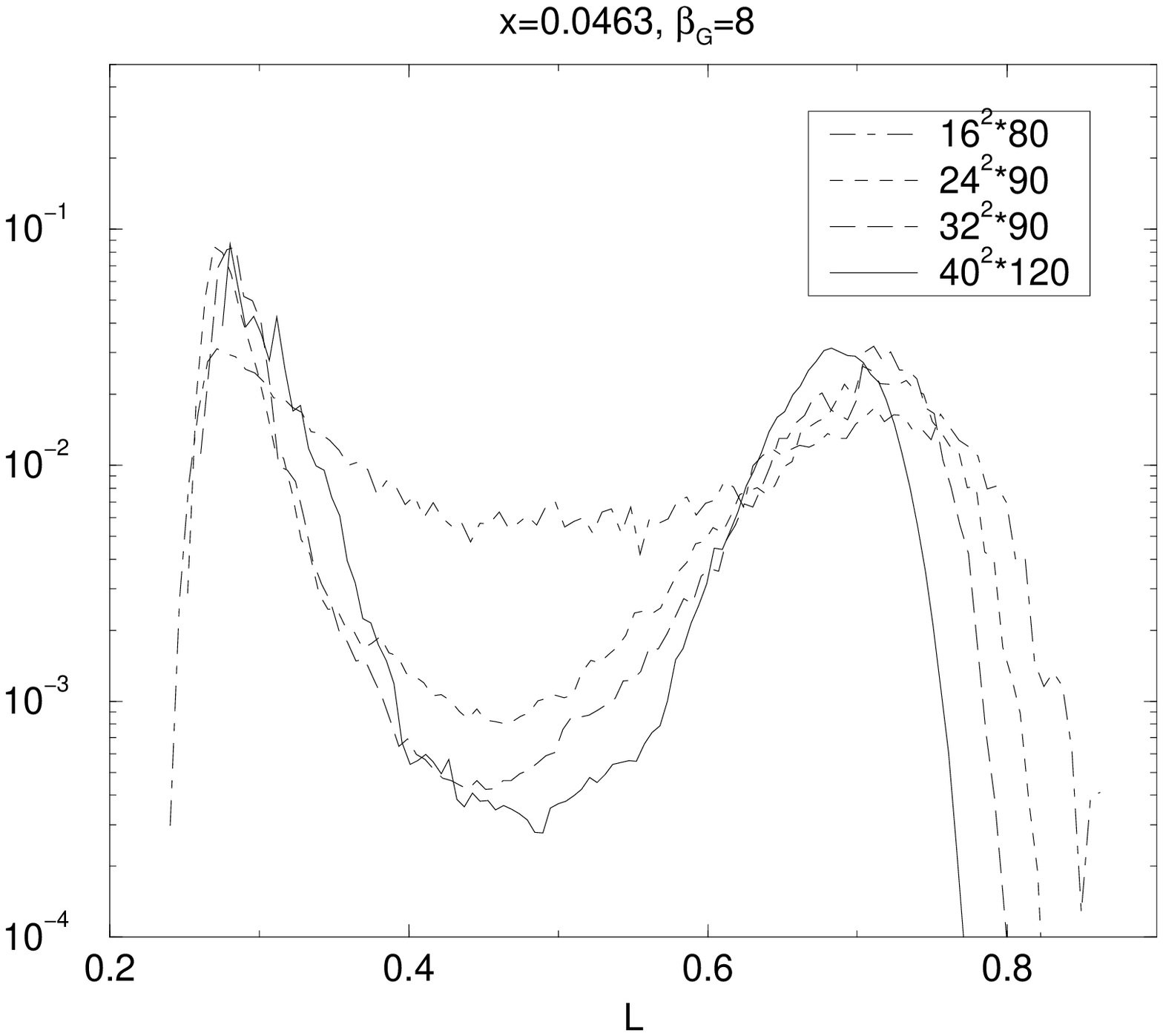}}
 
\vspace*{-0.3cm}

\centerline{\hspace{-2mm}
\epsfxsize=6.5cm\epsfbox{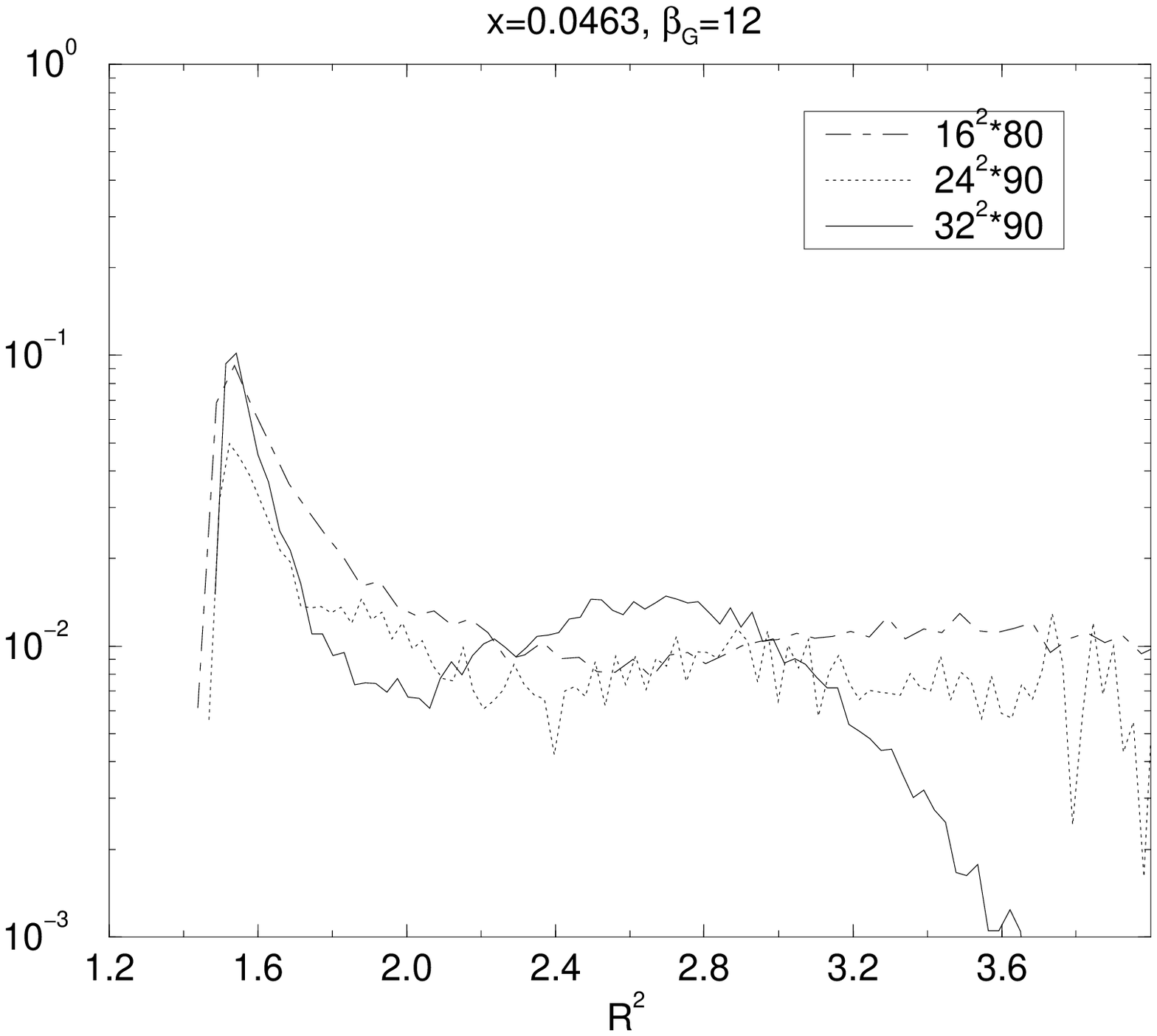}
\hspace{0cm}
\epsfxsize=6.5cm\epsfbox{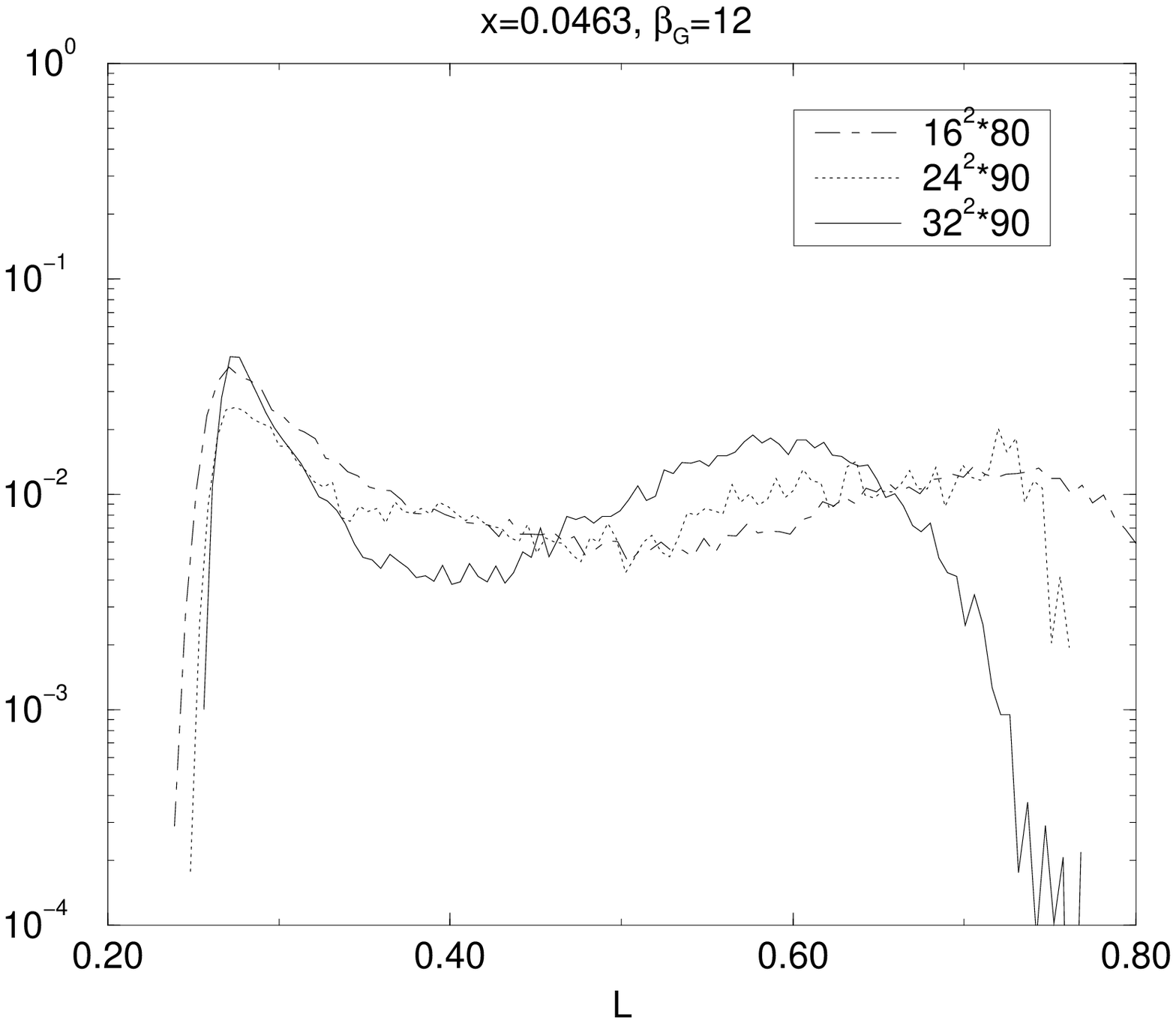}}
 
\vspace*{-0.5cm}

\caption[]{Equal weight histograms of $R^2$ and $L$ for $x=0.0463, 
\beta_G=4,6,8,12$. As the volume is increased the peaks remain at fixed
positions as expected for a 1st order transition. For $\beta_G=12$
a proper $V\to\infty$ limit would necessitate larger lattices.}
\label{figure:eqweightbg4}
\end{figure}


The equal weight histograms for $\beta_G=4,6,8,12$ are shown in
Fig.~\ref{figure:eqweightbg4}. 
The histograms are obtained by joining
all data together and then reweighting it to obtain the point where the
areas of the two peaks are equal. This involves choosing a value to
separate the
two peaks. The choice is arbitrary, and introduces some systematic
error to the determination of the critical point. We have used the
same value for all lattices for a given $x, \beta_G$-pair. This value
was determined by the minimum of the distribution in the largest
volume. The error we quote is purely statistical: it was obtained by
a jackknife analysis.

\begin{figure}

\centerline{\hspace{-2mm}
\epsfxsize=8cm\epsfbox{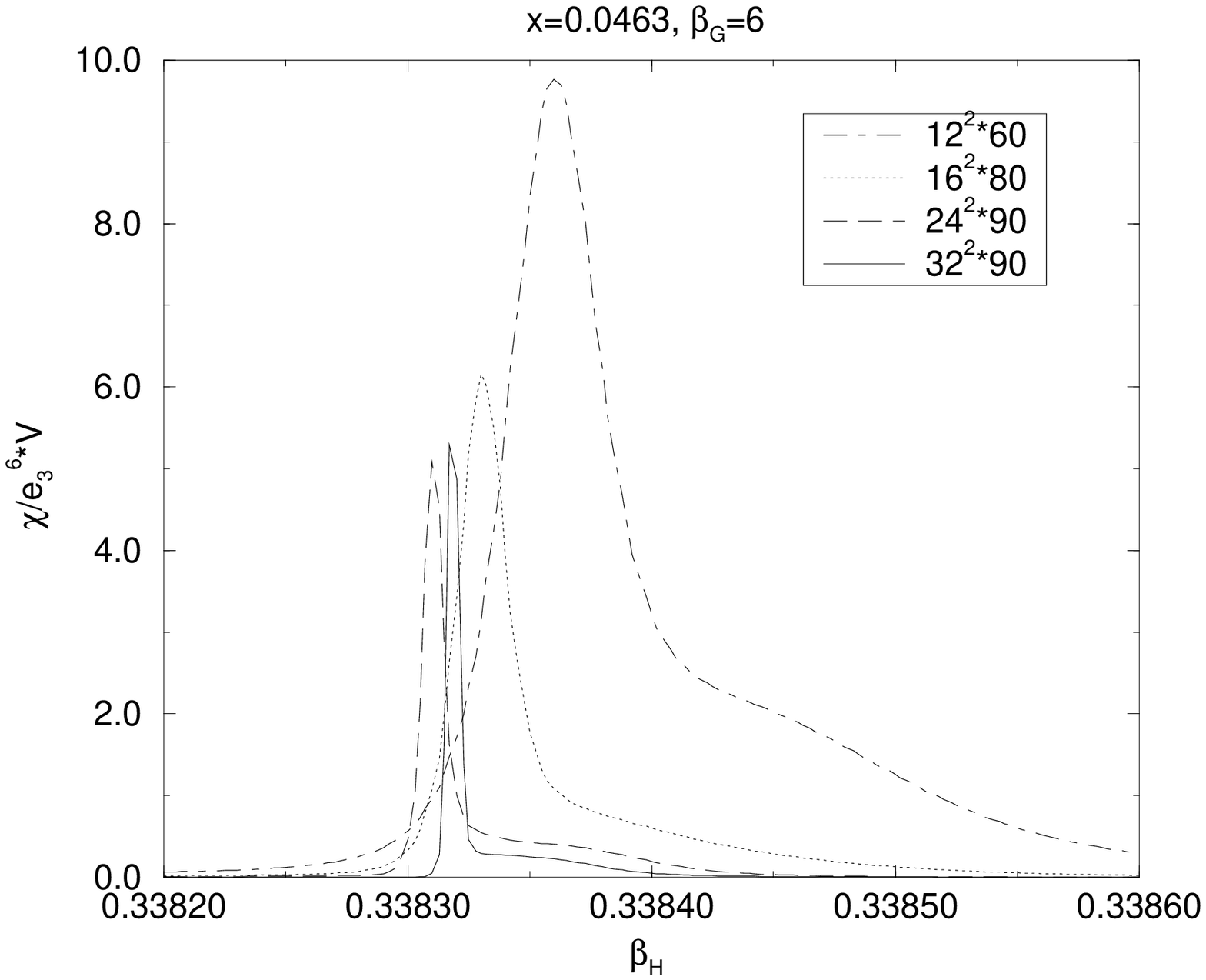}
\hspace{0cm}
\epsfxsize=8cm\epsfbox{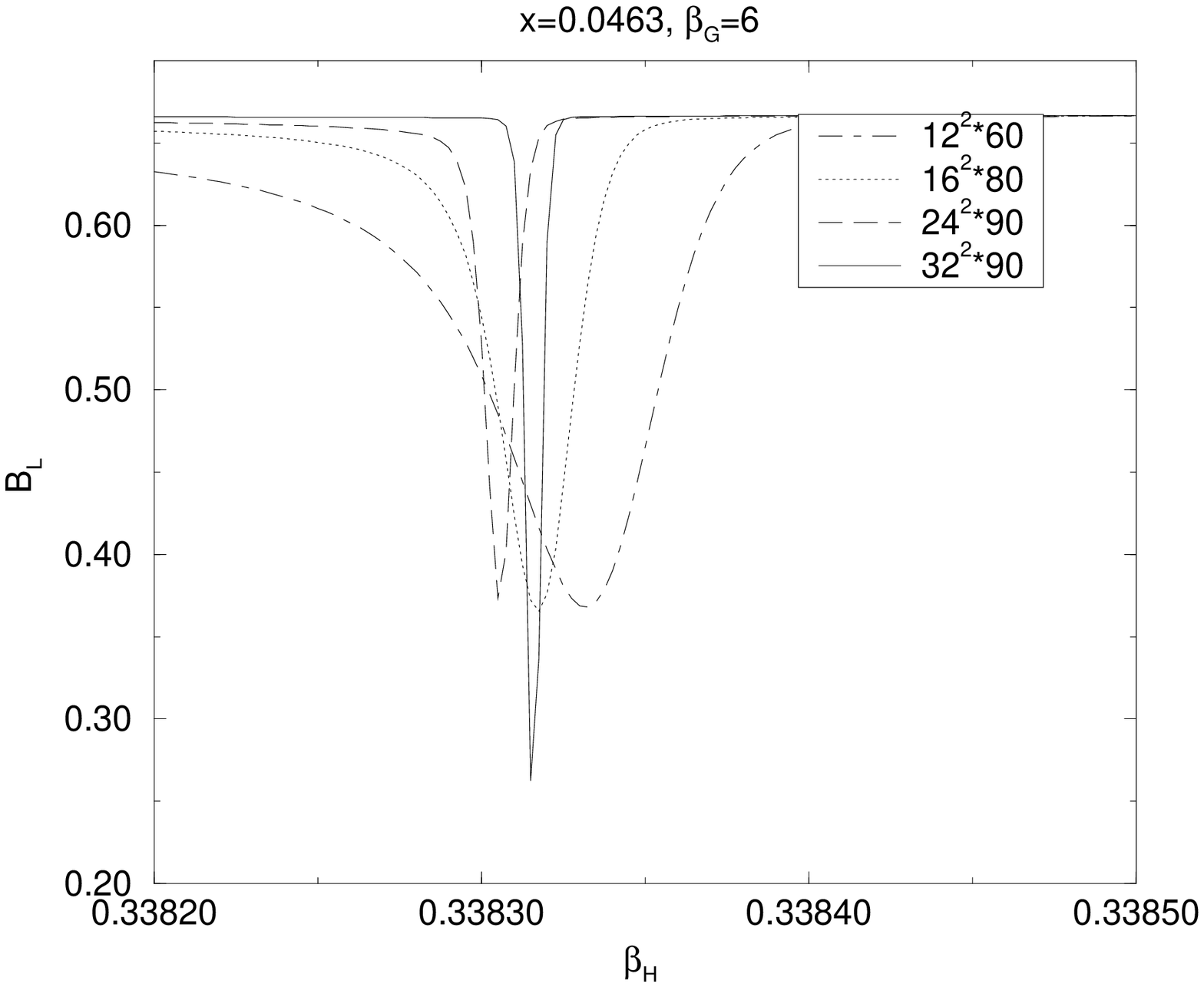}}

\vspace*{-0.5cm}

\caption[]{The susceptibility 
of $R^2$ divided by the volume $V$ (left) and the
Binder cumulant of $L$ (right) at $x=0.0463, \beta_G=6$.}
\label{figure:suskis}
\end{figure}

The locations of the maxima of the two susceptibilities and the minimum
of the cumulant were also obtained by joining all data and then
reweighting each jackknife block independently. As an example, the
susceptibility of $R^2$ and the Binder cumulant of $L$ for $\beta_G=6$
are plotted in Fig.~\ref{figure:suskis}. The maxima of $\chi_{R^2}$
are plotted in Fig.~\ref{figure:maxsuskis}.

\begin{figure}[t]

\vspace*{-1cm}

\centerline{\hspace{-2mm}
\epsfxsize=8cm\epsfbox{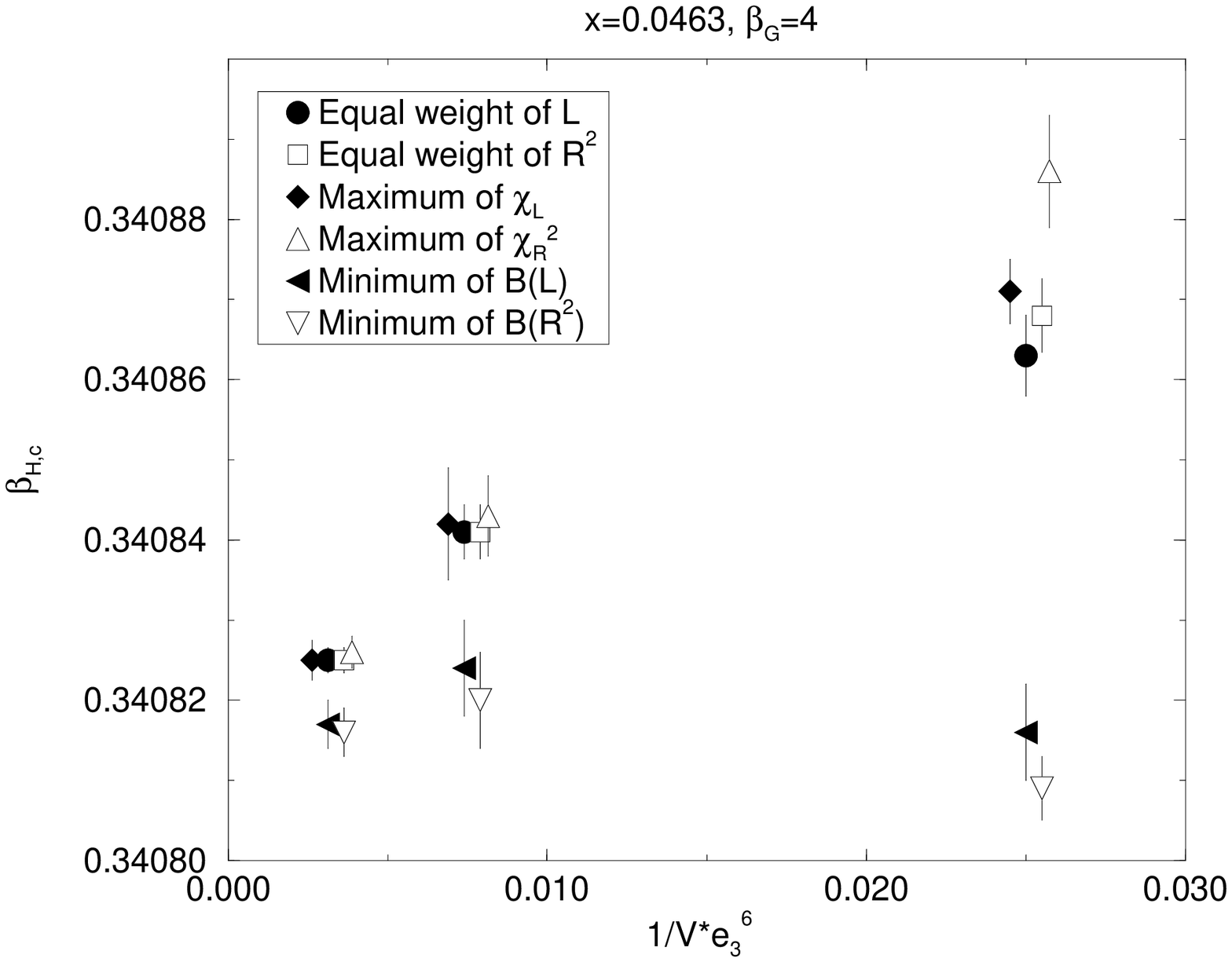}
\hspace{0cm}
\epsfxsize=8cm\epsfbox{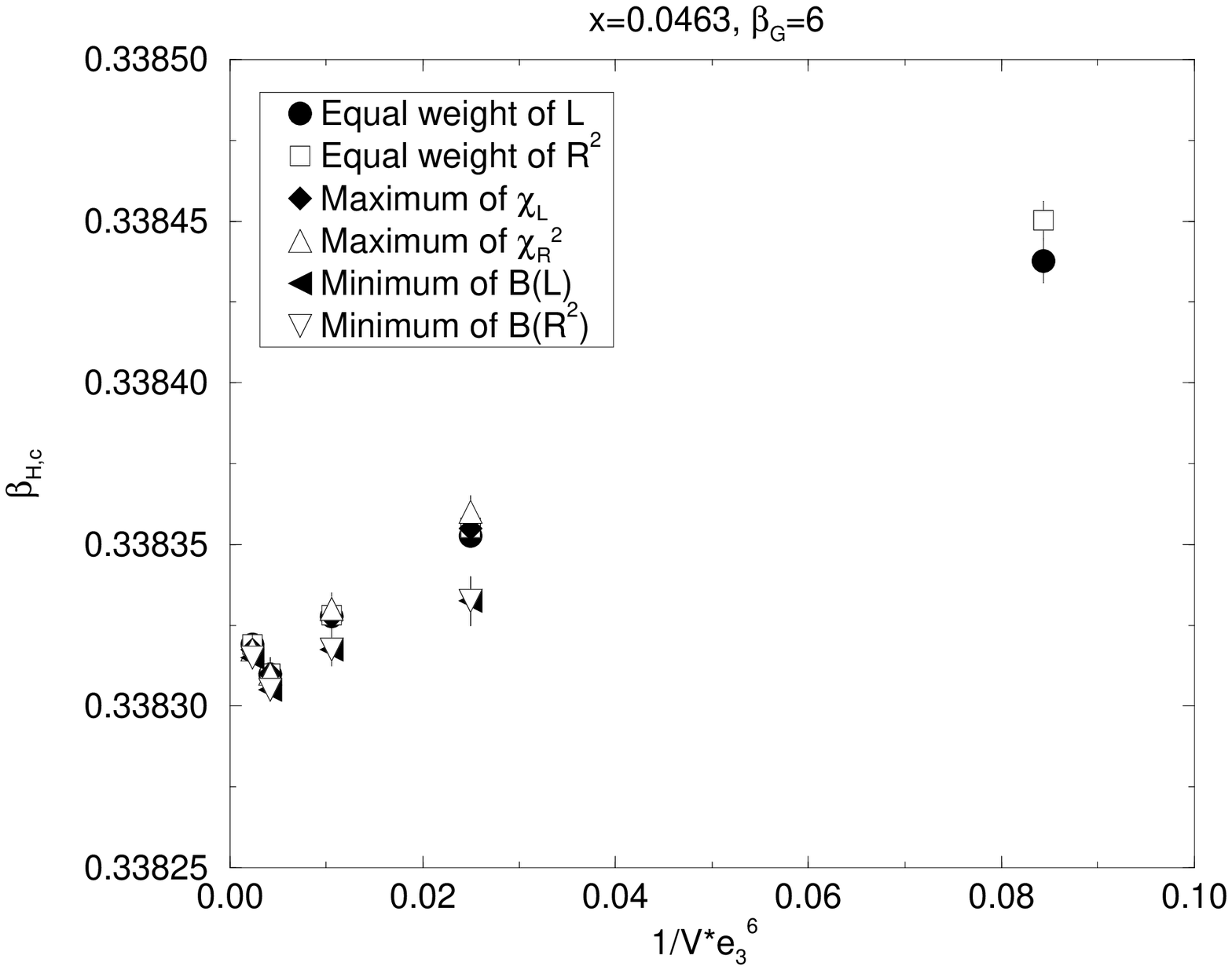}}
 
\vspace*{-0.5cm}

\centerline{\hspace{-2mm}
\epsfxsize=8cm\epsfbox{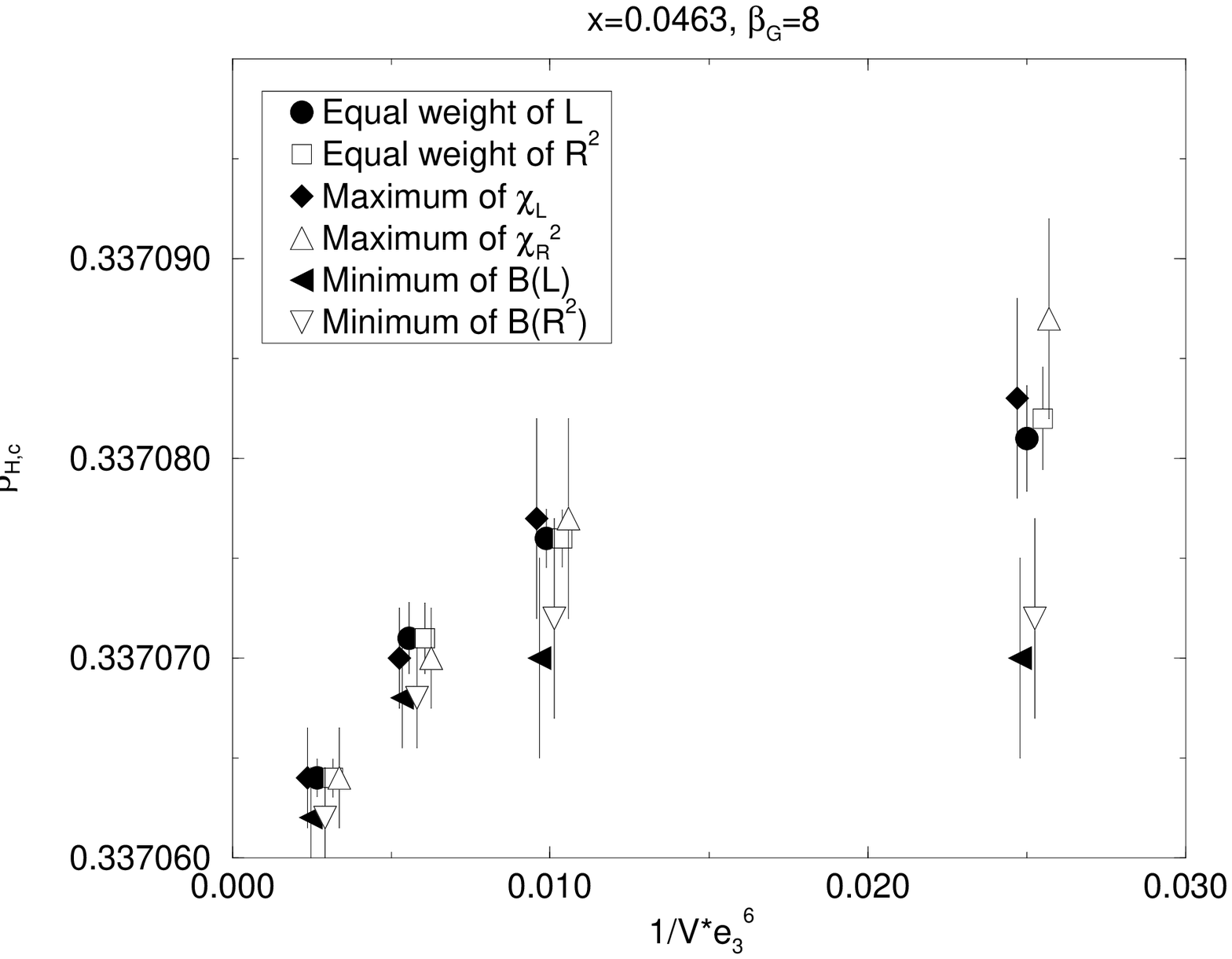}
\hspace{0cm}
\epsfxsize=8cm\epsfbox{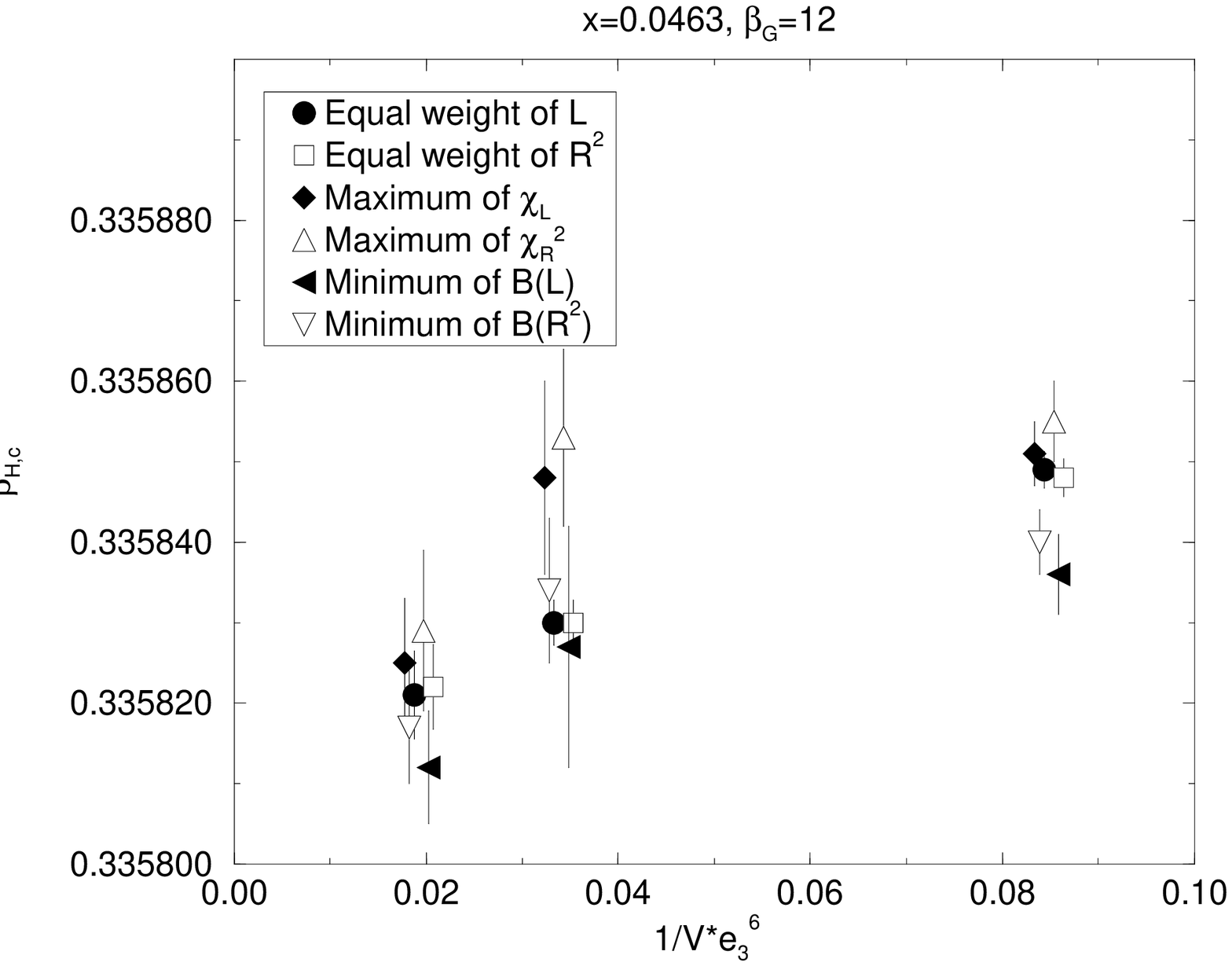}}

\vspace*{-0.5cm}

\caption[]{Different pseudocritical points 
for $x=0.0463, \beta_G=4,6,8,12$ as
a function of the volume.}
\label{figure:infvol.bg12}
\end{figure}

In Fig.~
\ref{figure:infvol.bg12} we have
plotted different pseudocritical 
points for $x=0.0463, \beta_G=4\ldots 12$.  The values obtained from
different 
methods differ at a finite volume, but as one increases the
volume these pseudocritical values come closer to each other
and agree with each other at the infinite volume limit (within statistical
errors). The results obtained from different methods are not 
statistically independent, but serve as a consistency check for the
infinite volume extrapolation. The different pseudocritical points at
the infinite volume limit are
collected in Table~\ref{table:yc}.

\begin{table}[ht]
\center
\begin{tabular}{|c|c|c|c|c|c|} \hline
$\beta_G$ & $p(R^2)$ & $p(L)$ & $\max
\chi_R^2$ & $\max \chi_L$ & $\min B_L$ \\ \cline{1-6}
4 & .3408199(18) & .3408208(19) & .3408208(25) & .3408192(28) &
.3408189(34) \\
6 & .3383148(07) & .3383151(06) &  .3383073(37) & .3383111(24) &
.3383075(23) \\
8 & .3370630(10) & .3370630(10) & .3370629(16) & .3370639(23) &
.3370638(23)  \\
12 & .3358169(40) & .3358165(40) & .3358314(107) & .3358235(92) & 
.3358069(86) \\ \cline{1-6} 
\end{tabular}
\caption{The critical coupling $\beta_{H,c}$ 
extrapolated to infinite volume for all
lattice spacings. The first two columns refer to the equal weight
criterion.} 
\label{table:yc}
\end{table}

\begin{figure}[t]
\centering
\leavevmode
\epsfxsize=8cm
\epsfbox{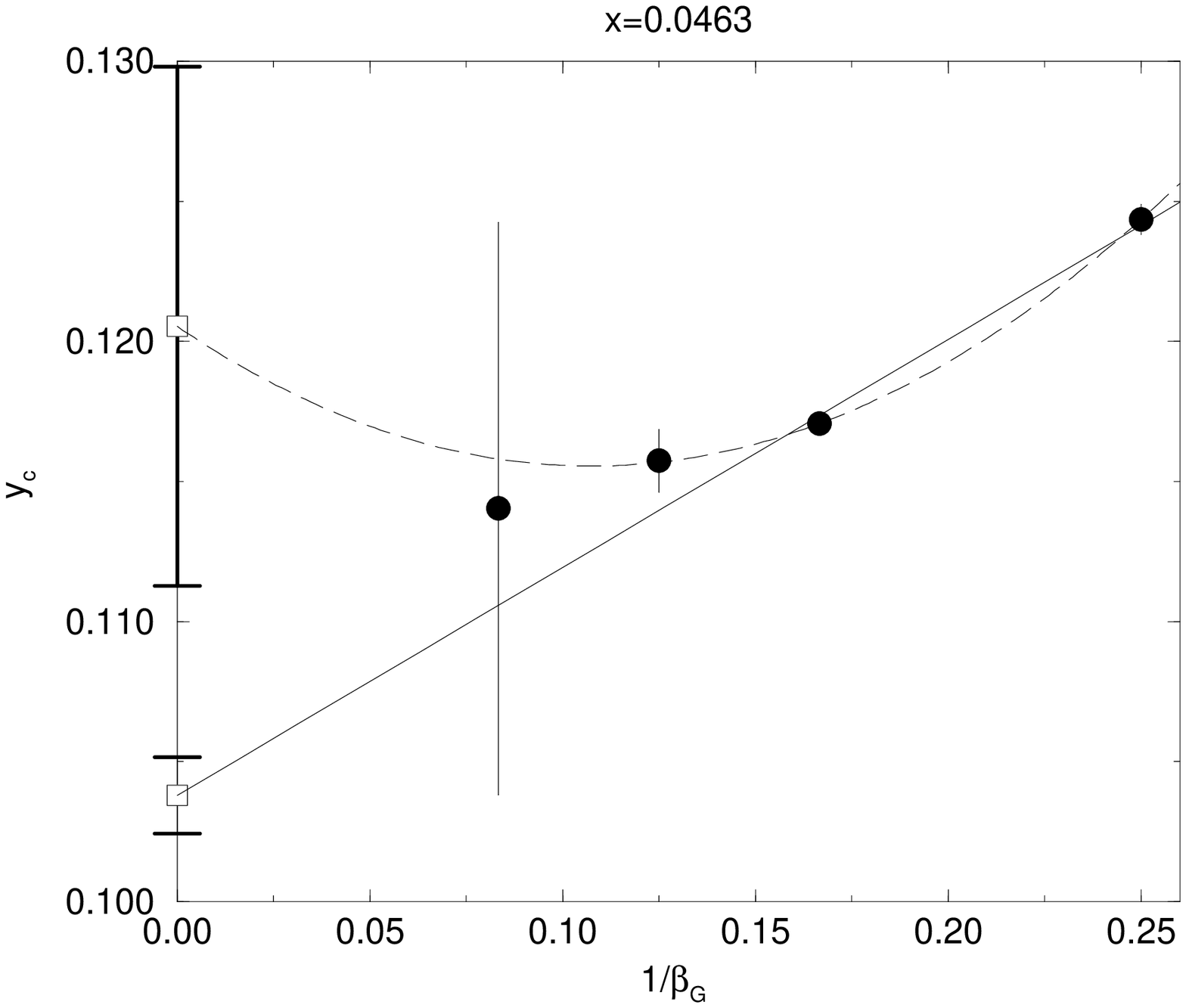}

\vspace*{-0.5cm}

\caption{The infinite volume extrapolations of the
pseudocritical points as a function of the lattice
spacing $ae_3^2=1/\beta_G$. Quadratic and linear fits 
for the limit $a\to0$ are also shown. }
\label{figure:contlimit}
\end{figure}

The extrapolation of $y_c$ to
continuum is shown in Fig.~\ref{figure:contlimit}.
A linear extrapolation has a confidence level of $19\%$ and gives
the value  $y_c = 0.1038(14)$. The quadratic fit gives
$y_c=0.120(9)$. As the linear fit is reasonably good, we choose
to quote that value as our continuum result. 
The 2-loop perturbative value $y_c=0.0947$ is more than $5\sigma$ from
this.

It is possible to compare results at $\beta_G=4$ and $8$ 
to those obtained with
non-compact simulations in~\cite{dfk}. 
We find that our critical couplings
$\beta_{H,c}$ do not agree with the
critical couplings obtained with noncompact simulations.
{}From Fig.~5 in~\cite{dfk} one can read that the critical value
for $\beta_G=4$ is roughly $0.34040$, with a very small error (smaller
than $10^{-5}$). At $\beta_G=4$ we obtain $\beta_{H,c}=0.3408208(19)$,
so the statistical error cannot explain this difference. This is not
unexpected, as there is no need for the
non-compact and compact results
to agree at a finite lattice spacing. 
However, in the continuum limit one
should obtain the same results. 
The authors of \cite{dfk} choose to quote 
the critical temperature instead of
the critical coupling $y_c$. If one extrapolates the results from
\cite{dfk} to continuum with a linear fit, one obtains $T_c =
130.86(58)$. Using their eq.~(13) one can try to compare our
continuum value with this non-compact result. Our linear fit
result corresponds to $T_c = 131.28(4)$, so that
the two results agree well in the continuum limit.

\subsubsection{Type II: $x=2$}

\begin{figure}[t]
\centering
\leavevmode
\epsfxsize=8cm
\epsfbox{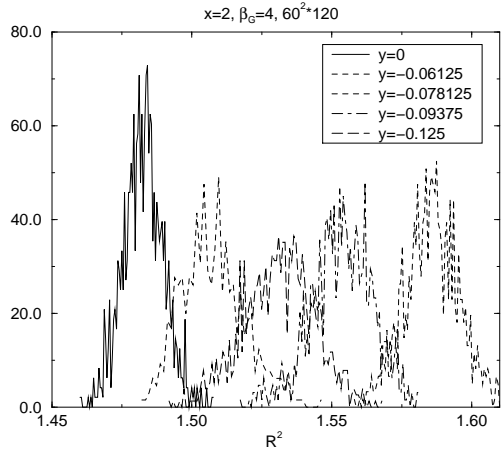}

\vspace*{-0.5cm}

\caption{Histograms for $R^2$ at $x=2$,  
around the critical point $y\approx -0.05$.
No two-peak structure is seen.}
\label{figure:hgramsx2}
\end{figure}

The location of the transition point in the type II regime has 
already been 
discussed in~\cite{u1prb}. 
Measurements of averages of local operators 
in the type II regime show no evidence of discontinuities.
Typical distributions are shown in Fig.~\ref{figure:hgramsx2}.
To quantify the effect, the maxima of the $R^2$ susceptibility 
are shown in Fig.~\ref{figure:maxsuskis} as a function
of the volume. One can see that 
the maximum of $\chi_{R^2}$ remains constant. This
finite size scaling behaviour is expected if there is no transition or 
if the transition is of
second order with a critical exponent $\alpha \le 0$. For understanding
the type II regime and for quoting a definite value for $y_c$, 
mass measurements, especially that of $m_\gamma$,
are required.

\begin{figure}
\centering
\leavevmode
\epsfxsize=8cm
\epsfbox{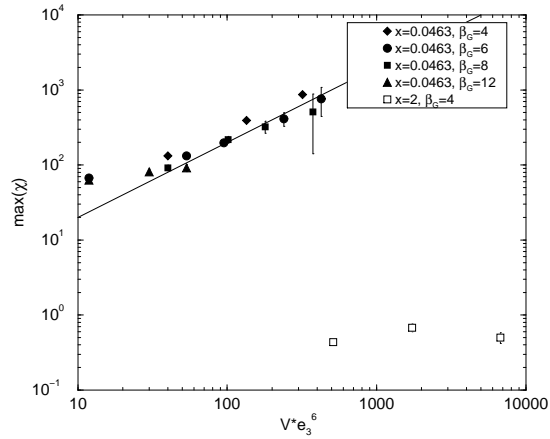}

\vspace*{-0.5cm}

\caption{The maximum of the susceptibility of $R^2$ for $x=0.0463$ 
and $x=2$ as a function of the volume.}
\label{figure:maxsuskis}
\end{figure}

\subsection{Latent heat}

\begin{figure}

\centerline{\hspace{-2mm}
\epsfxsize=8cm\epsfbox{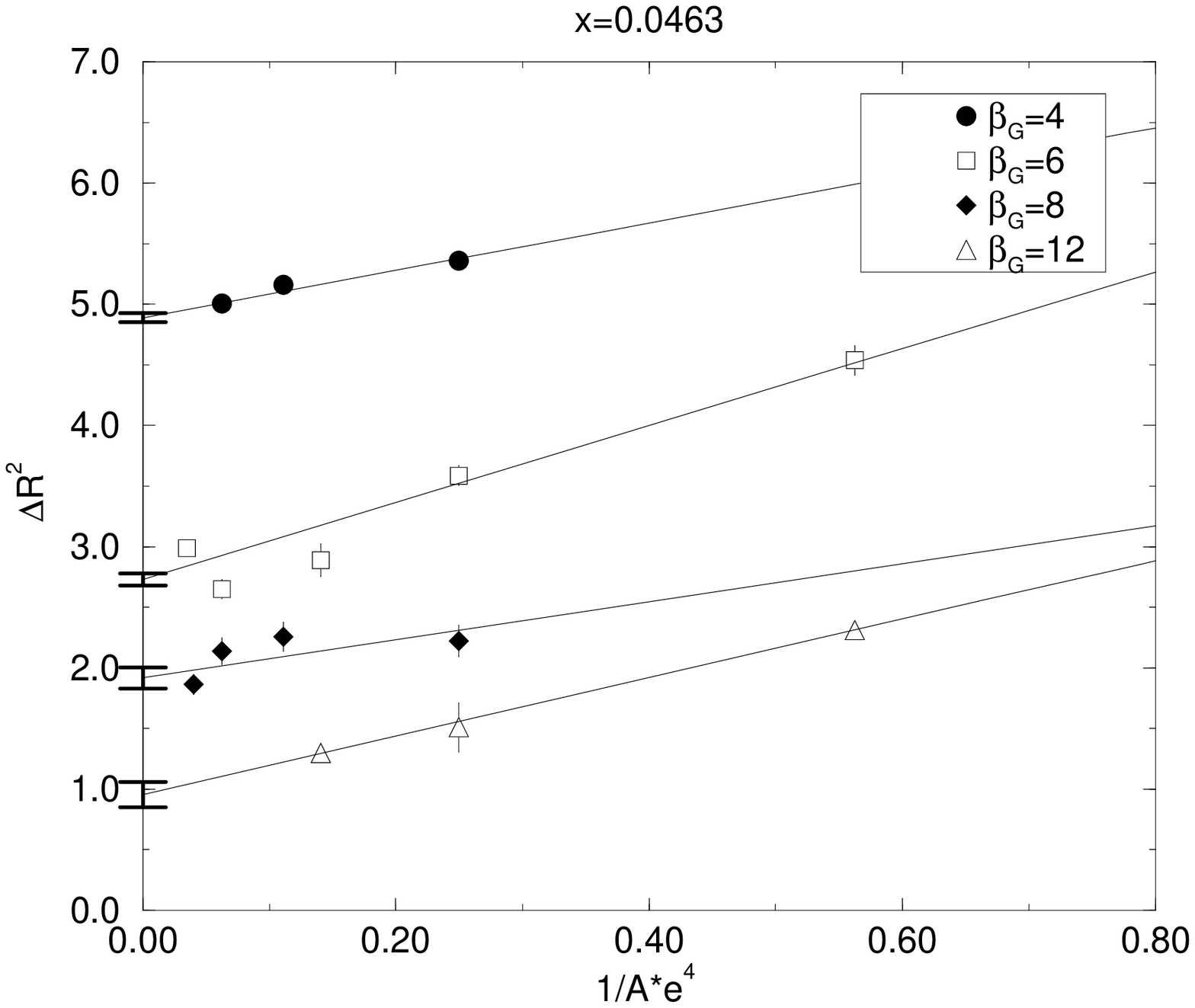}
\hspace{0cm}
\epsfxsize=8cm\epsfbox{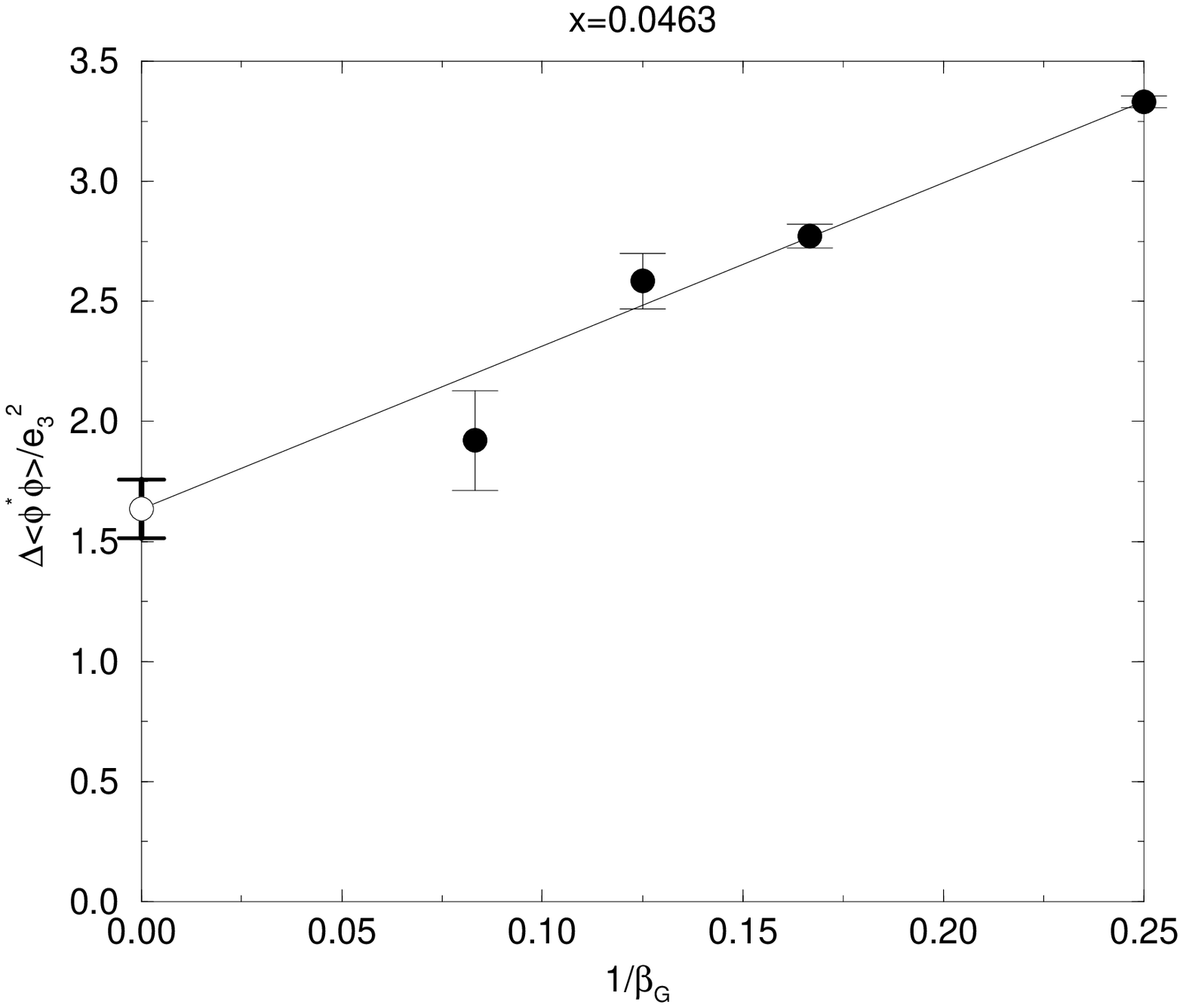}}

\vspace*{-0.5cm}

\caption[]{The infinite volume limits of $\Delta R^2$ (left)
and the continuum 
limit of $\Delta\ell_3$ (right).}
\label{figure:latentheat}
\end{figure}

The latent heat --- or the energy released in the transition ---
is directly proportional to $\Delta\ell_3 = \Delta\left<\phi^*\phi\right>$
(see, e.g., Sec.~11 in \cite{nonpert}). This quantity is easily computed
non-perturbatively from lattice simulations. In
Fig.~\ref{figure:latentheat} we have 
plotted the infinite volume extrapolations of $\Delta\left< R^2
\right>$. These can be converted to continuum values by
eq.~(\ref{scaling}), 
\be
{\Delta \left< \phi^* \phi \right>\over e_3^2} = {1\over 2} \beta_H
\beta_G \Delta \left< R^2 \right>.
\ee
The infinite volume limit is taken by extrapolating linearly with 
the inverse
area of the system, as extrapolation with the inverse volume would fail to
accommodate the small differences between the different ratios of
$N_x/N_z$. 

The continuum limit of $\Delta\ell_3$ is displayed
in Fig.~\ref{figure:latentheat}. The continuum extrapolation is done
with a linear fit, which seems to fit the data well -- the confidence level
of this fit is $28\%$. The final continuum value we obtain is 
$\Delta\ell_3/e_3^2 = 1.64(12)$, with a statistical error only. 

Again the results should be compared with those obtained from
perturbation theory and from non-compact lattice simulations. In
\cite{dfk} it was found that the value of $\Delta\ell_3$ increases with
decreasing lattice spacing; however we find a rather strong
decrease. Therefore, even though all our data except $\beta_G=12$ give
a value higher than the perturbative one, our continuum value is
considerably smaller than the perturbative value $\Delta\ell_3=2.25$.
Taking into account the fact that our data at $\beta_G=12$ is of not as
high quality as at smaller lattice spacings by using only the three
largest lattice spacings for the continuum extrapolation, one still obtains
a value smaller than the perturbative one, $\Delta\ell_3=1.71(13)$. 

\subsection{Interface tension}

\begin{figure}

\centerline{\hspace{-2mm}
\epsfxsize=8cm\epsfbox{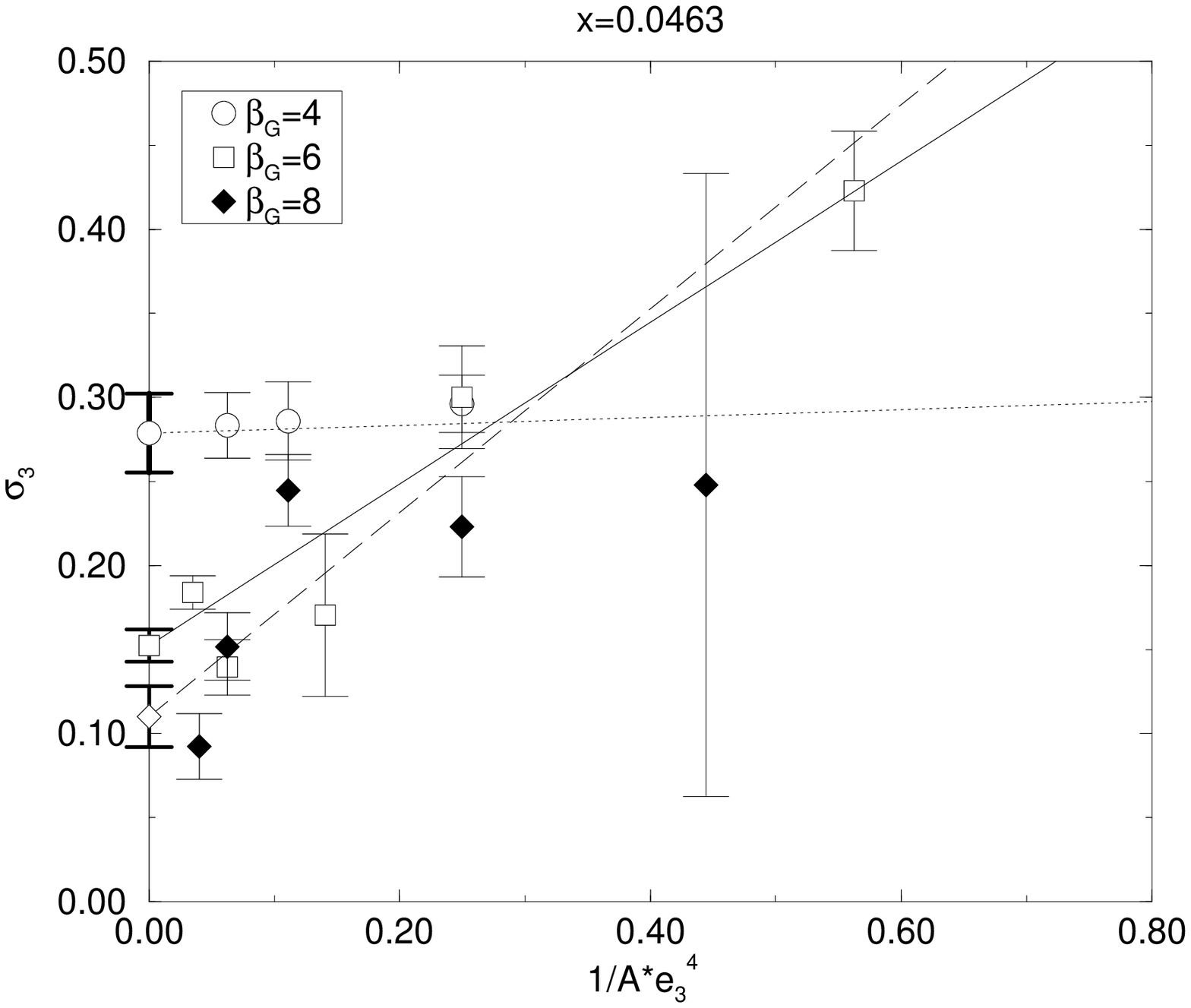}
\hspace{0cm}
\epsfxsize=8cm\epsfbox{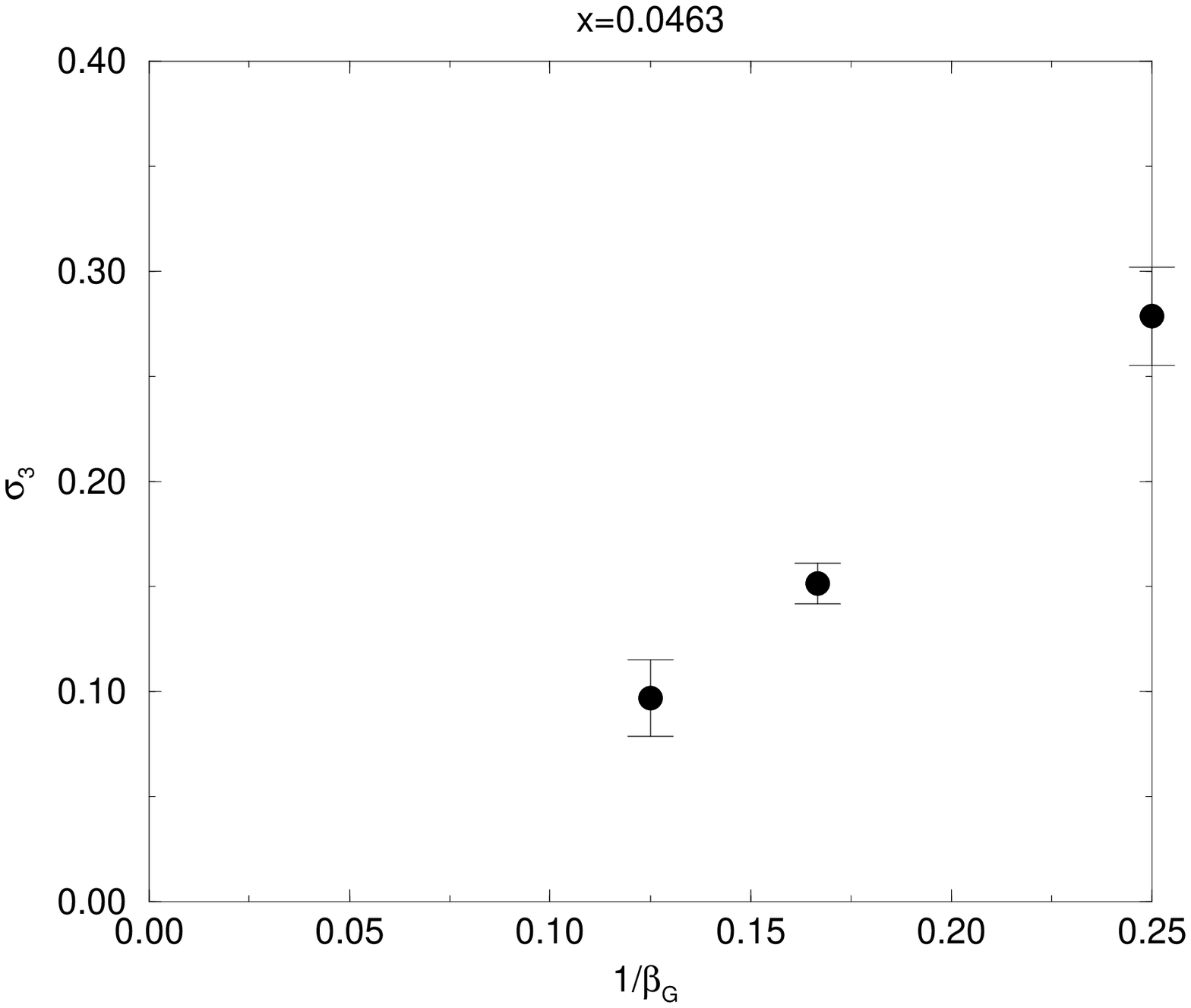}}

\vspace*{-0.5cm}

\caption[]{The infinite volume limit (left) and the 
extrapolated results at different
lattice spacings (right) for the dimensionless interface tension $\sigma_3$.
The systematic errors are expected to be much larger than the statistical
ones shown in the figures.} 
\label{figure:sigma}
\end{figure}

The interface tension is the most difficult of the characteristics
of the first order phase transition to measure. 
We estimate it with the histogram method \cite{binder}: at
the critical point the
system can reside in a mixed state consisting of domains
of the pure phases. Because some extra free energy is associated with
the interface separating the two phases, the area of the interface tends to
a minimum and in a system with a geometry $N_x^2\times N_z$ (with $N_z
> N_x$), the minimum area is just $2\times N_x^2a^2$ (due to
periodic boundary conditions at least two interfaces are needed). 
Since the
cost of an interface is $\sim \exp[-\sigma A/T]$,
the dimensionless interface tension $\sigma_3$ can be calculated from
\be
\sigma_3 =  {1\over 2 A e_3^4}\log{P_{max}\over
P_{min}}, 
\label{sigma}
\ee
where $P_{min}$ and $P_{max}$ are the minimum and the maximum of the
probability distribution, and
\be
e_3^4 A = e_3^4 N_x^2 a^2 = \left({N_x\over \beta_G}\right)^2.
\ee
For finite size corrections in 
the interface tension, see, e.g., \cite{ikkry}.

The formula \nr{sigma} requires that the two interfaces are
far enough from each other so that their mutual interaction can be
neglected. This would be signalled by a flat minimum in order
parameter histograms, which should appear at large enough values of
$N_z$. As seen from 
Fig.~\ref{figure:eqweightbg4}, 
a clearly flat minimum is not yet observed: the interfaces are so thick
that their interaction is not negligible even for lattices of length 90.
Thus non-negligible systematic errors are expected in our results. 

The infinite volume extrapolations and the values obtained at different
lattice spacings are shown in Figure~\ref{figure:sigma}. 
No extrapolation to continuum limit is performed even though the data
seems to be consistent with a linear fit. 
This is since a linear fit would
give a negative value for $\sigma_3$, which is 
unacceptable. 
We ascribe this unphysical behaviour to larger systematic
errors at larger $\beta_G$'s, where the physical volume is 
not large enough to contain regions of clearly separated phases.
The results we obtain are $\sigma_3 = 0.28(2)$ at
$\beta_G=4$, $\sigma_3 = 0.15(1)$ at $\beta_G=6$, and $\sigma_3 =
0.09(2)$ at $\beta_G=8$; $\beta_G=12$
does not give a good enough signal 
for this purpose.  

The perturbative value for $\sigma_3$ is 0.225, and as the results at
$\beta_G=6$ and $8$ are both below this and the value seems to be
decreasing with decreasing lattice spacing, one expects that
the real value is significantly lower than the perturbative one. 
In Fig.~\ref{Dl3} we have shown the conservative estimate
$\sigma_3=0.14(14)$ which contains all the values we have measured. 

\subsection{Mass measurements}

Since bulk quantities are rather insensitive to the phase
transition at large values of $x$, we have paid special attention to
mass (= inverse correlation length) measurements. 
Most of the results of our measurements have been reported in~\cite{u1prb}, 
and here we describe the techniques used in some more detail. There are 
several factors which could result in
too high a value for a mass. First, one has to be sure that there is no
contamination from higher excited states. This can be ensured by 
systematically searching for optimized operators coupling to 
the desired excitations, using blocking and diagonalization 
techniques. One also has to start the
fits from sufficiently large values of $z$ and to monitor the
behaviour of the effective mass. Second, one should use large
enough lattices --- in particular, one cannot expect to measure reliably
masses smaller than $~\sim 1/N$.

\subsubsection{Blocking}

\begin{figure}
\centering
\leavevmode
\epsfxsize=8cm
\epsfbox{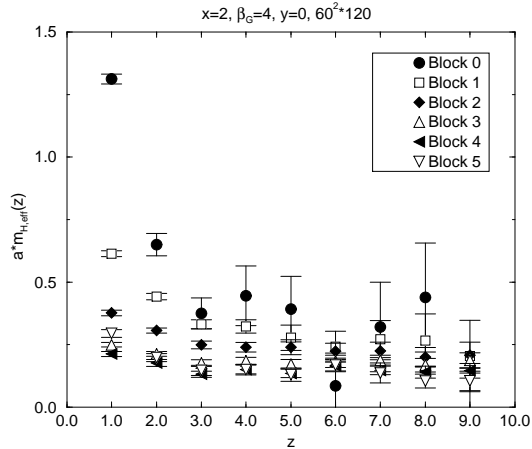}

\vspace*{-0.5cm}

\caption{Effective masses at different blocking levels at $x=2,
\beta_G=4$ as a function of $z$ (in lattice units).}
\label{figure:blocking}
\end{figure}

To avoid having to use extremely large starting values of $z$ it is
necessary to have a very good projection on the ground state. 
The projection can be improved upon 
by using blocking
in the transverse direction~\cite{phtw, teper} 
both for the link variables $U_i(x)$,
\be
U_i^{n+1}(x)
= \fr13\biggl[
U^n_i(x)U^n_i(x+\hat{i}) + \sum_{j \ne i}
U^n_j(x)U^n_i(x+\hat{j})U^n_i(x+\hat{j}+
\hat{i})U^n_j(x+2\hat{i})^{-1}\biggr],
\ee
and for the scalar fields $\phi(x)$,
\be
\phi^{n+1}(x) = 
\fr15\biggl[\phi^n(x)+
\sum_{i=\pm 1}^{\pm 2} U^n_i(x)\phi^n(x+\hat{i})\biggr].
\ee
The operators measured are then constructed from the 
blocked fields according to eqs.~\nr{oo}--\nr{oi}.
The iteration level $n$ can be tuned for optimal overlap.
In Fig.~\ref{figure:blocking} we show the effect of
blocking. Typically we find that the best results are obtained at
blocking level 3 for all operators. The overlap of our blocked
operators with the
ground state ranged from being consistent with 100\%
far away from the critical point to 75\% at the critical point.

\subsubsection{Variational analysis}

The blocking described above is 
in practice sufficient for the scalar excitations,
where we are mostly interested in the ground state. However, we would
like to know not only the ground state of the vector excitation but
also have some information on the excited state. We are especially
interested in knowing to what extent the lowest vector excitation
couples to the operator $\tilde V_3$, which consists purely of gauge
fields. To study this we have performed a variational analysis,
which enables us to measure also the excited state. 

Variational analysis is based on the fact that we have several
operators which couple to the 
same quantum numbers. Even if each individual
operator is not perfect in the sense that it also contains
contributions from higher excited states, it should be
possible to reduce this
effect by suitably choosing linear combinations of individual
operators. The basis that we use consists of different operators
measured at different blocking levels. To be specific we use blocking
levels 2 and 3; for scalar excitations we use the operators $\phi^*\phi$ and
$\mbox{Re}\,\phi^*U_i\phi$ and for vector excitations
$\mbox{Im}\,\phi^*U_i\phi$ and $\sin\hat F_{jk}$, 
see eqs.~\nr{oo}--\nr{oi}. Thus in both cases we
have four different operators, which we call $O_i$. We then measure
the whole $4\times 4$ cross correlation matrix
\be
C_{ij}(z) = \left< O_i^*(z) O_j(0) \right>,
\ee
and then construct the improved operators ${\bf O_i}$
in the standard way~\cite{varanal}, 
\be
{\bf O_i} = \sum_i v^i_k O_k.
\ee
The operators are further normalised to unity at zero distance.
The coefficients 
of the operator ${\bf O}_1$ for the ground state 
are found by maximising 
\be
\left< {\bf O_1}^*(a){\bf O_1}(0)\right>.
\label{maxvar}
\ee
The operators for excited states can then be obtained by maximising the
analogy of
eq.~(\ref{maxvar}) in a subspace that is orthogonal to the ground
state operator ${\bf O}_1$.  

We find that the variational analysis has
usually little effect for the ground
state --- however, for excited states it is extremely helpful.

\subsubsection{Fits}

We extract the different masses by fitting exponentials to the corresponding
correlation functions. In practice we have used one and two
exponential fits and find that the difference between these two is within
statistical errors. However, the one exponential fit is usually more
stable, and therefore all numbers we quote are from one
exponential fits.

One should note that the
expectation values of the correlation function $C(z)$ are strongly
correlated for close values of $z$. Typically the statistical
correlation between $C(z)$ and $C(z+1)$ is greater than 95\%.
Therefore it is necessary to take this correlation into account when
performing the fits. We find that the effect of taking this into
account is to constrain the fits more tightly. The $\chi^2$ is
increased and the variation between different jackknife blocks is
decreased. If we do not use correlation information in our fits the
fit parameters can fluctuate more freely causing not only
unrealistically large errors but also changing the final value of the
fit. 

The major systematic error comes from the choice of the fitting
range. We have carefully monitored the behaviour of the local mass,
and start the fits only when a clear plateau is visible. Still changing
the starting point by one or two lattice spacings changes the result;
however as the change typically is roughly equal to the statistical error,
we do not quote a systematic error on our fits. When interpreting the
results one should bear in mind that it is possible that the
systematic errors are at least as large as the statistical ones.

\subsubsection{Polyakov mass}

\begin{figure}
\centering
\leavevmode
\epsfxsize=8cm
\epsfbox{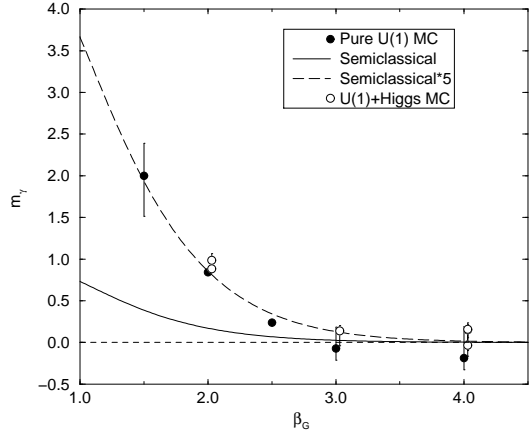}

\vspace*{-0.5cm}

\caption{The photon mass as a function of $\beta_G$ in 
compact 3d U(1) gauge
theory. The 5 $\times$ semiclassical curve extrapolates between weak and
strong coupling regimes, and reproduces the non-perturbative
values reasonably well for intermediate $\beta_G$.}
\label{figure:polyakov}
\end{figure}

Let us apply the methods described first to the 
photon mass in the compact formulation, see eqs.~\nr{pmass}, 
\nr{pmassstrong}.
In order to use $m_\gamma$ as an order parameter, one has to be sure
that this Polyakov mass it acquires at a finite lattice spacing, 
is negligible for all practical purposes. 
The semiclassical calculation (\ref{pmass}) gives 
$m_\gamma/e_3^2 = 0.0033$ at $\beta_G=4$, which should be much smaller
than anything we expect to be able to measure. However, as most of our
conclusions depend strongly on $m_\gamma$, we have checked the
validity of the semiclassical calculation by direct simulations in the
pure U(1) gauge theory. We used a $24^3$ lattice, varied
$\beta_G=1\ldots 4$ and used 
the operator \nr{oi} to measure $m_\gamma$.

The Monte Carlo results for pure gauge theory are shown in
Fig.~\ref{figure:polyakov}, together with the semiclassical result
\nr{pmass} and
three data points in the U(1)+Higgs theory. The U(1)+Higgs data points
were taken in the symmetric phase: we chose $x=2, y=0$ and
$y=0.5$. One sees that adding the Higgs field does not affect the
Polyakov mass, and that the Monte Carlo data 
agree with the semiclassical result for $\beta_G \gsim 3$ 
(in~\cite{ws}, agreement with semiclassical computations was found even 
at $\beta_G\sim 2$, but this relies on an ``improved''
$\beta_G$. Using that, the agreement in Fig.~\ref{figure:polyakov}
is improved, as well). At $\beta_G = 4$
the mass is clearly too small to be seen, and therefore can be
neglected for all practical purposes. For lower values of
$\beta_G$ we find that the formula $5\;\times$ semiclassical result
reproduces the Monte Carlo data reasonably well, effectively
interpolating between the strong and weak coupling results 
in eqs.~\nr{pmass}, \nr{pmassstrong}.

\subsubsection{Masses}

\begin{figure}

\centerline{\hspace{-2mm}
\epsfxsize=8cm\epsfbox{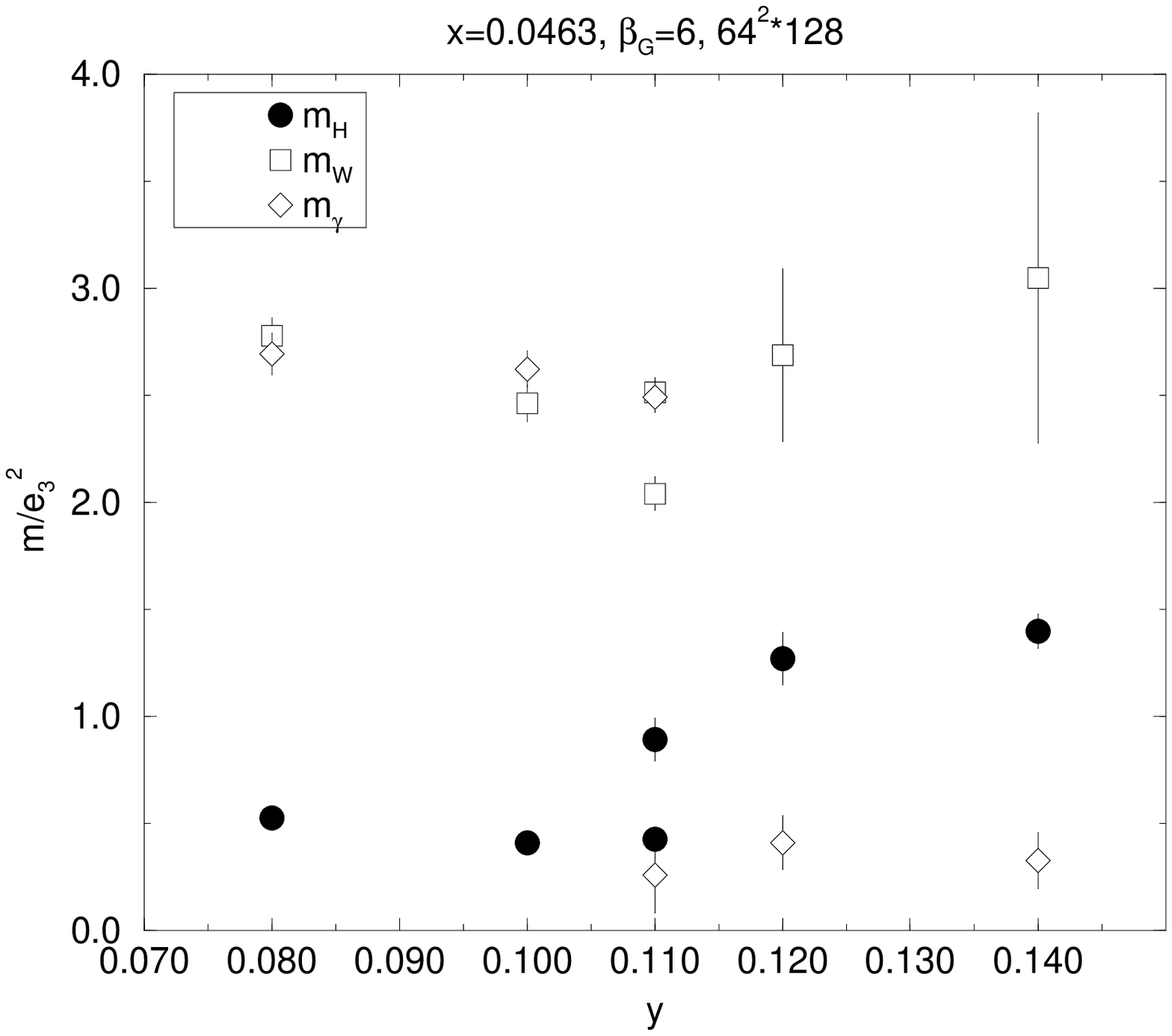}
\hspace{0cm}
\epsfxsize=8cm\epsfbox{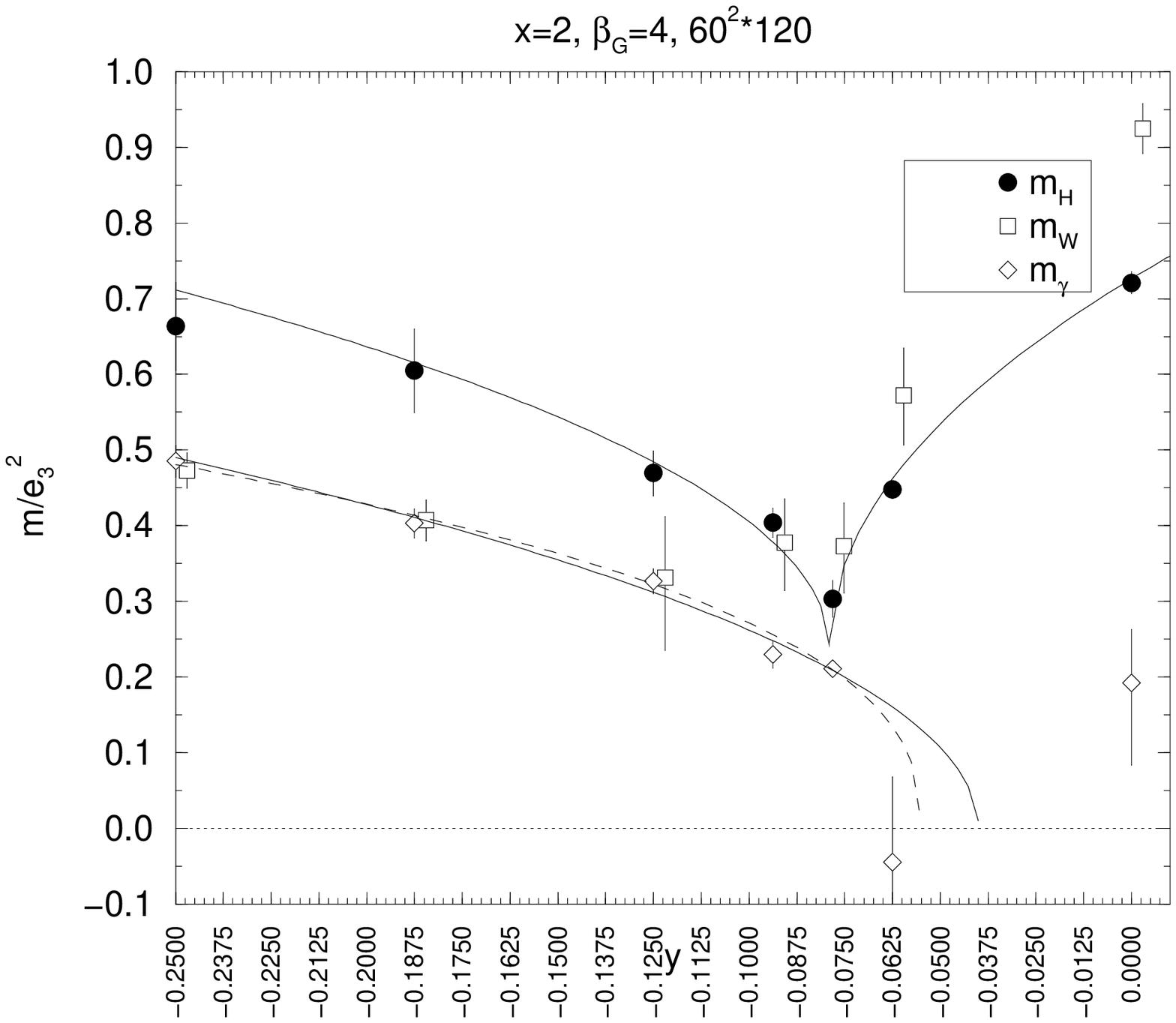}}

\vspace*{-0.5cm}

\caption[]{Masses near the critical point at $x=0.0463, \beta_G=6$
(left) and $x=2, \beta_G=4$ (right) (from~\cite{u1prb}).}
\label{figure:masses}
\end{figure}

The masses 
in the U(1)+Higgs theory 
obtained with the methods described above are plotted in
Fig.~\ref{figure:masses}. We see that the mass of the photon
vanishes within
$2\sigma$ for both $x=0.0463$ and $x=2$ in the symmetric phase.
When measured in $\sigma$'s, some of
the results at $x=0.0463$ are somewhat further from zero. We believe
that this due to the fact that no variational analysis was done for
these points. Thus $m_\gamma$ serves as an order parameter
for practical purposes.

The first order nature of the transition at $x=0.0463$
is clearly visible. All the masses show a clear discontinuity at
$y_c(x)$. When going from the broken to the symmetric phase, 
$m_H$ jumps upwards (the scalar correlation length decreases from
about 12$a$ to about 7$a$). The two vector excitations $m_\gamma$
and $m_W$, which were degenerate for $y<y_c$, separate so that
$m_\gamma$ jumps to zero and $m_W$ becomes the mass of an excited
two-scalar bound state. 

At $x=2$ the pattern
is less clear. The mass of the scalar excitation
$m_H$ dips deeply, but does not go to zero. 
The value of $m_H$ at the minimum was also shown to be independent
of the volume within statistical errors~\cite{u1prb}. 
In the broken phase
the two vector operators again see one single excitation, but
in the symmetric phase $m_\gamma$ has dropped to zero and the other
excitation has a large mass. However, on the basis of this data
one cannot conclusively state what is happening. One can envisage
the following possibilities:
\bi
\item There is one line $y_c(x)$ on which $m_\gamma$ goes continuously
to zero, $m_H$ has a minimum and $m_\gamma, m_W$ separate.
\item As the previous one but $m_\gamma$ jumps discontinuously to zero.
This does not imply that the transition is of first order, as there is no
jump in local quantities.
\item The line $y_c(x)$ has split into several branches.
\item All the masses go to zero at $y_c(x)$. This is the standard
2nd order scenario, which however, is not supported by our data.
\ei

The data shown is at fixed $\beta_G=4$ and one may ask whether
the extrapolation to continuum would change the conclusions. 
The minimum of $m_H$ in Fig.~\ref{figure:masses} corresponds to
a correlation length of 16$a$ and it is hard to see how making $a$
still smaller could change the results.
We have made some runs at $\beta_G=3,6$ and did indeed not find 
any dependence outside of statistical errors.

\subsection{Critical exponents}

If $m_\gamma$ goes to zero continuously, one can measure its
critical exponents.
If the transition were of second order in the usual
sense, there would be critical exponents for both scalar and vector
correlation lengths:
\ba
m_H &\sim& |y-y_c|^{\nu_H}, \\
m_\gamma &\sim& |y-yc|^{\nu}.
\ea
However, as we do not see any critical behaviour
for the scalar correlation length, we cannot measure any critical
exponent corresponding to it.
The approximate scaling of $m_H$ away from the critical 
point has been discussed in~\cite{u1prb}.

To measure 
the critical exponent $\nu$ corresponding to a diverging photon
correlation length, we have fitted functions of the form
$m_\gamma/e_3^2 = A(y_c-y)^\nu$ to the data shown in
Fig.~\ref{figure:masses}, both
keeping the critical exponent $\nu$ fixed to the mean field value $0.5$ or
allowing it to vary freely. We have also varied the range of $y$ which
we include 
in the fits as $y_{min} = -1.25\ldots -0.25$. The fits with their
confidence levels are shown in Table~\ref{table:critfits}. We find that the
critical point is rather stable, and that the value of the critical
exponent is consistent with the mean field value. All the fits have rather
low confidence levels, but as noted previously the errors we quote for
masses are statistical only. Inclusion of the systematical errors
(which we estimate to be of the same order of magnitude
as the statistical errors) should
bring the confidence levels to perfectly acceptable values. 

In principle one could also measure the anomalous dimension $\eta_N$
of the photon correlation length, 
$\left<\tilde V_3(\bfx)\tilde V_3(0)\right> \sim
1/|\bfx|^{1+\eta_N}$, but this would require a significantly extended
analysis and more accurate data. 

\begin{table}
\center
\newcommand{\fx}{$^*$}
\begin{tabular}{|c|c|c|c|c|}
\cline{1-5}
$y$ range & $A$ & $y_c$ & $\nu$ & CL \\
\cline{1-5}
$-0.25\ldots 0$ & 1.07(04) & -0.040(05) & 0.5\fx & 0.14 \\
$-0.25\ldots 0$ & 0.91(24) & -0.055(23) & 0.39(17) & 0.10 \\
$-0.5\ldots 0.5$ & 0.995(19) & -0.033(04) & 0.5\fx & 0.08 \\
$-0.5\ldots 0.5$ & 0.931(62) & -0.048(13) & 0.429(62) & 0.07 \\
$-1.25\ldots 0.5$ & 0.959(11) & -0.026(04) & 0.5\fx & 0.006 \\
$-1.25\ldots 0.5$ & 0.947(12) & -0.046(06) & 0.440(19) & 0.06 \\
\cline{1-5}
\end{tabular}
\caption[]{\label{table:critfits}
Fits to determine the critical exponent $\nu$. 
The superscript (\fx)
after a value means that it was kept fixed.}
\end{table}

\section{Conclusions}

We have in this paper studied the 3d U(1)+Higgs model
as an effective theory of various finite temperature theories.
There are two separate aspects of the problem: computing the
parameters of the effective theory in terms of the physical
parameters of the full theory and studying the 3d theory as such.
The first aspect has been solved for a very concrete physical phenomenon,
superconductivity (Appendix~A), and for a more academic system, hot scalar 
electrodynamics (Appendix~B). 

The second aspect, 
the study of the 3d theory as such, also splits into two
parts: the region of first order transitions, 
corresponding to small $x=\lambda_3/e_3^2$
(type I superconductors, or small Higgs masses), and
the region of possibly second order transitions, large $x$ 
(type II superconductors, or large $>m_W$ Higgs masses). 
The type II regime has been analysed in~\cite{u1prb}, 
and here we have mainly concentrated on the type I regime. 

The evidence for a first
order phase transition at $x=0.0463$ is compelling: several order
parameter like quantities display a clear discontinuity, the
finite volume scaling of the susceptibities is clearly consistent with
a first order transition and the mass of the photon can be used as an
effective order parameter. However, we have found that even when 
the transition is relatively strong, there are quantitatively
some discrepancies with perturbation theory which seems to predict 
too strong a transition, see Fig.~\ref{Dl3}.
This is in contrast to the case of the SU(2)+fundamental Higgs 
model, for example, where 2-loop perturbation theory works quite
well for comparative Higgs masses~\cite{nonpert}.
The reason is probably just that the transition is weaker
in U(1)+Higgs, but one might also speculate that physical
topological defects (vortices) may play a role. 
In any case, the transition is too weak to be observed to 
be of first order in practical superconductor experiments.

At $x=2$ the situation is even more subtle. Many of the operators that
can be used to study the phase transition at $x=0.0463$ do not display
any indication of a transition at $x=2$. None of the local parameters ($R^2$
and $L$) show any discontinuity. The maxima of the susceptibilities
remain constant. The scalar correlation length increases, but still
seems to remain finite when extrapolated to 
infinite volume and zero lattice spacing
\cite{u1prb}. The only indication of the transition is a 
vector correlation length diverging within statistical errors. 

One thing to be checked
in this argument is the use of the compact formulation
which, in fact, makes the photon massive
at a finite lattice spacing, if only by an
exponentially small amount.  
Thus in a strict sense
the transition vanishes altogether
for a finite $a$. 
We have therefore
studied the topological
effects related to the compact formulation and demonstrated that 
they behave as expected, vanishing 
rapidly when approaching the continuum limit. Thus 
we believe that they do not alter 
the pattern described above.

The precise properties of the transition at $x=2$ remain unclear. Our data
cannot distinguish between a second order phase transition in which
the mass of the photon vanishes continuously and a more exotic 
scenario in which the photon mass has a discontinuity. To solve
this unambiguously one would clearly need much more data.
Another study requiring even more data would be that 
of the endpoint $x_c$ of the first order regime.

Because of the non-abelian gauge structure 
one might have expected computations in 3d SU(2)+Higgs 
theories~[10,22--28] 
to be more complicated than those in 
U(1)+Higgs theories. However, just the opposite is the case. U(1)+Higgs
is more demanding to simulate than SU(2)+Higgs at least
in the compact formulation, as some of the correlation
lengths are larger, the transition is weaker, the structure of the phase
diagram is more complicated, and very large lattices are needed. This is
analoguous, say, to what happens in the $q$-state Potts model
in which the strength of the transition rapidly increases with the
number $q$ of field components. Perhaps also the formation of 
physical topological defects, {\em viz.}\ vortices,
plays a role in U(1)+Higgs.

\section*{Acknowledgements}
 
We acknowledge useful discussions 
with B. Bergerhoff, J. Jers\'ak, S. Khlebnikov, C. Michael,
O. Philipsen, K. Rummukainen, M. Shaposhnikov, M. Tsypin and G. Volovik.

\appendix
\renewcommand{\thesection}{Appendix~~\Alph{section}:}
\renewcommand{\thesubsection}{\Alph{section}.\arabic{subsection}}
\renewcommand{\theequation}{\Alph{section}.\arabic{equation}}

\section{\la{FtI}Full theory I: BCS superconductivity and the 
Ginzburg-Landau model}

So far we have only treated the 3d effective theory as such; 
in these two Appendices we
shall quantitatively discuss how two different physical theories
map to the same effective theory. The first case is
superconductivity. 

Quantum phenomena in superconductors are microscopically described by the 
BCS theory \cite{bcs}. In the normal state electrons do not form bound 
coherent states because of the repulsive Coulomb force but at low 
temperatures the electrons can form Cooper pairs due to their interaction
with the ionic lattice. The electron-phonon interaction is important only at 
temperatures where the thermal excitations from the Fermi energy of the 
electron 
gas have an energy smaller
than the average phonon energy. When this condition is 
satisfied the Fermi sea becomes unstable against the formation of 
bound pairs of electrons from states above the Fermi surface.  

When the BCS theory is treated within the framework of the Green's 
functions method one can find the connection between the microscopic 
BCS theory and the effective macroscopic Ginzburg-Landau model of 
superconductivity.
Using the electron-electron interaction of the BCS theory and 
solving the equations of motion of the Hamiltonian Gor'kov 
\cite{gorkov} derived an equation describing the free energy 
density of Cooper pairs (we keep here $c,\hbar$):
\ba
f({\bfx})&=&N(0)\bigg[r\xi_0^2
\left|\left(\nabla -i \frac{2e}{\hbar c}A_i\right)\Delta({\bfx})\right|^2+
\nonumber\\
&&\quad+\left(\frac{T}{T_0}-1\right)|\Delta({\bfx})|^2 +
\frac{7\zeta(3)}{8(\pi T_c)^2}|\Delta({\bfx})|^4 \bigg], \label{go}
\ea  
where $\Delta({\bfx})=\langle \psi_{\uparrow}({\bfx})
\psi_{\downarrow}({\bfx})\rangle$, $r\le1$ (=1 for perfect crystals)
is an impurity parameter of the material,
\be
\xi_0=\sqrt{\frac{7\zeta(3)}{48}}\frac{\hbar v_F}{\pi T_c}
\ee
sets the typical length scale of the system, and 
\be
N(0)=\frac{m^2 v_F}{2\pi^2\hbar^3}
\ee
is the density of states at the Fermi surface.
Note that the temperature parameter $T_0$ equals
$T_c$ only on the mean field level.
The connection between the free energy density \nr{go} and the 
parameters in eq.~\nr{parameters} of
the U(1)+Higgs theory can now be established
by scaling the fields so that \nr{go} has the form 
in eq.~\nr{ac}.
The final relation becomes
\be
    y=\frac{1}{rq^4}\left(\frac{T}{T_0}-1\right), \,\,\,\, x=\frac{g}{(rq)^2},
\la{scparameters}
\ee
where we have used the notations of \cite{kleinert}:
\ba
    g=\frac{3 T_c}{N(0)\hbar^2v_F^2\xi_0}=
\sqrt{\frac{108 \pi^6}{7 \zeta(3)}} \left(\frac{T_c}{T_F}\right)^2, 
\nonumber \\
q=\frac{2e}{\hbar c}\sqrt{T_c \xi_0}=\sqrt{\frac{4 e^2}{\hbar c \pi} 
\sqrt{\frac{7\zeta(3)}{48}}}\sqrt{v_F\over c},
\ea   
with $T_F=mv_F^2/2 \gg T_c$. What now is crucial is that widely
different scales with large or small ratios appear.
For low temperature superconductors ($T_c\sim 1$ K), 
$T_c/T_F \sim 10^{-4}$ and the Fermi velocity is $v_F \sim 
(10^{-3} - 10^{-2}) c$. These lead to values $g\sim10^{-6}$ 
and $q\sim 0.01$ resulting in $x\sim 0.01/r^2$. For usual 
superconductors thus $x\ll1$. In contrast, for high temperature 
superconductors typically $x\gg 1$. This follows from measurements
of the penetration depth
and the coherence length and from the fact that $x$ is
simply half the square of their ratio, so that $x<1/2$ describes type I 
superconductors and $x>1/2$ type II superconductors. It is interesting
that the physically challenging case (high $T_c$ superconductors)
just corresponds to the region in which the G-L model has a
particularly subtle structure. It is so in the relativistic case
(Appendix B), as well, that the large $x$ regime correponds to the 
most challenging 
domain of the physical theory: large Higgs masses.

The big dimensionless ratios also appear in the 
expression for $y$ in eq.~\nr{scparameters}. 
This implies that although $y_c(x)$ is of order one
for small $x$, $T_c/T_0-1$ is extremely small, of the order of
$q^{-4}\sim 10^{-8}$. This smallness follows not from the dynamics
of the 3d theory, but from the relation of the full and effective
theories.

Similar statements apply in the case of the latent heat $L$. In the
effective theory, $L$ is essentially the jump  $\Delta\ell_3$ in 
$\langle\phi^*\phi\rangle$ at the phase transition point. 
For small $x$ (see eq.~\nr{leading};
now again $\hbar=c=1$),
\be
 \Delta \ell_3=\frac{1}{18 \pi^2 x^2}
\approx\frac{r^4 q^4}{18 \pi^2 g^2}. \la{1lDl3}
\ee 
The conversion into physical units \cite{nonpert} gives then
\be
    L=\Delta \ell_3 e_p^6 T_c^5\frac{dy}{dT}=
 T_c^4 \frac{r^3 e_p^6}{18 \pi^2 g^2} \simeq 1.15 \times 10^{-18} 
K_c^4 \frac{r^3 e_p^6}{18 \pi^2 g^2} \mbox{J}/\mbox{cm}^3,
\ee 
where $e_p=2e=\sqrt{16 \pi \alpha} \simeq 0.61$ is the coupling 
constant of the physical theory and $K_c$ is the transition 
temperature in Kelvins.
For example, for Pb ($K_c=7.2$ K) the latent heat is
\be
L \simeq 10^{-6} \mbox{J}/\mbox{cm}^3 = 10^{-5} \mbox{J}/\mbox{mole},
\la{mole}
\ee
which is far too small to be detected experimentally at present.
This is so in spite of the fact that the transition is quite strong
from the point of view of the effective theory.
Note also that non-perturbatively the transition is  
somewhat weaker than given by eq.~\nr{1lDl3}, see Fig.~\ref{Dl3}, 
but this does not change the order of magnitude estimate in eq.~\nr{mole}.

\section{\la{FtII}Full theory II: 
hot relativistic scalar + fermion electrodynamics}

Secondly, we shall work out the 3d effective theory of finite 
$T$ relativistic
scalar electrodynamics. Since we aim at precision, we here first
have to discuss how the parameters of the Lagrangian are
determined to 1-loop in the $\msbar$ scheme in terms of the
measurable (in principle!) physical masses of the theory \cite{generic}.
Next we have to relate this 4d finite $T$ theory and the 3d effective
theory. This happens in two stages: 
first the momenta $p\gsim \pi T$ (the superheavy modes) 
are integrated out, resulting in an effective bosonic 3d theory with 
fundamental and adjoint Higgs fields. Finally, the adjoint Higgs field
(the heavy modes) is integrated out. Some parts of the latter
two steps have been 
already discussed in~\cite{perturbative,blaizot,andersen2}.
The parametric accuracies of the effective theories derived
have been discussed in~\cite{generic}.

\subsection{Relation between physics and the Lagrangian in $\msbar$
in 4d}\label{ropms}

The Lagrangian of 4d scalar + chiral fermion electrodynamics is~\cite{ae}
\ba
{\cal L} & = & 
\fr14 F_{\mu\nu}F_{\mu\nu}+(D_\mu\phi)^*(D_\mu\phi)
-\nu^2\phi^*\phi+\lambda(\phi^*\phi)^2 \nonumber \\
& & + \bar{\Psi}_L\slash\!\!\!\!{D}\Psi_L 
+ \bar{\Psi}_R\slash\!\!\!{\partial}\Psi_R
+ g_Y\bar{\Psi}(\phi^*a_L+\phi a_R)\Psi,
\label{4dl}
\ea
where 
\ba
& & F_{\mu\nu}  =  \partial_\mu A_\nu-\partial_\nu A_\mu,\quad
D_\mu\phi=(\partial_\mu+ieA_\mu)\phi,\quad
\slash\!\!\!\!{D}\Psi_L=\gamma_\mu(\partial_\mu+ieA_\mu)\Psi_L, \nonumber\\
& & \Psi_{R,L} =  a_{R,L}\Psi,\quad 
a_{R,L}=(1\pm\gamma_5)/2,\quad 
\phi=(\phi_1+i\phi_2)/\sqrt{2}.
\ea
At the tree-level, the couplings $\nu^2,\lambda,g_Y^2$  
are related to the physical masses $m_H$, $m_W$ and $m_t$ 
of the $\phi_1, A_\mu$, and $\Psi$ fields in the broken phase by  
\be
\nu^2=\frac{1}{2}m_H^2,\quad
\lambda=\frac{e^2}{2}
\frac{m_H^2}{m_W^2},\quad
g_Y^2=2e^2\frac{m_t^2}{m_W^2}. \label{0l}
\ee
The masses $m_H$, $m_W$, $m_t$ are analogous to the Higgs, W and top
masses of the Standard Model, and they can, in principle, be experimentally
determined. The gauge coupling $e^2$ is, in principle,  
determined in terms of some cross section. 

When one takes into account loop corrections, the theory 
requires renormalization, and the relations in eq.~\nr{0l} change.
Let us for definiteness regularize the theory in the $\msbar$ scheme
in $d=4-2\epsilon$ dimensions. At 1-loop level, 
the bare quantites are then related to the 
running renormalized quantities through
\ba
A_B & = & A(\mu)
\biggl[1+\frac{\mu^{-2\epsilon}}{16\pi^2\epsilon}\Bigl(
-e^2\frac{N_s+2 N_f}{6}
\Bigr)\biggr], \\
\phi_B & = & \phi(\mu)
\biggl[1+\frac{\mu^{-2\epsilon}}{16\pi^2\epsilon}\Bigl(
\frac{3e^2}{2}-\frac{g_Y^2}{2}
\Bigr)\biggr], \\
\Psi_B & = & \Psi(\mu)
\biggl[1+\frac{\mu^{-2\epsilon}}{16\pi^2\epsilon}\Bigl(
-\frac{g_Y^2}{4}
\Bigr)\biggr], \\
e_B^2 & = & e^2(\mu)
\biggl[1+\frac{\mu^{-2\epsilon}}{16\pi^2\epsilon}\Bigl(
e^2\frac{N_s+2 N_f}{3}
\Bigr)\biggr], \\
\nu_B^2 & = & \nu^2(\mu)
\biggl[1+\frac{\mu^{-2\epsilon}}{16\pi^2\epsilon}\Bigl(
-3e^2+4\lambda+g_Y^2
\Bigr)\biggr], \\
\lambda_B & = & \lambda(\mu)
+\frac{\mu^{-2\epsilon}}{16\pi^2\epsilon}\Bigl(
3e^4-6\lambda e^2+10\lambda^2+2\lambda g_Y^2-g_Y^4
\Bigr), \\
g_{Y,B}^2 & = & g_Y^2(\mu)
\biggl[1+\frac{\mu^{-2\epsilon}}{16\pi^2\epsilon}\Bigl(
2g_Y^2-3e^2
\Bigr)\biggr]. 
\ea
Here $N_s=1$ is the number of scalar fields and $N_f=1$ is 
the number of fermion fields; $N_s$ and $N_f$ are used just 
to show the origin of the different contributions.
The renormalized parameters of the theory run 
at 1-loop level as
\ba
\mu\frac{d}{d\mu}e^2(\mu) & = & 
\frac{\mu^{-2\epsilon}}{8\pi^2} \frac{e^4}{3}(N_s+2 N_f), \la{run1}\\
\mu\frac{d}{d\mu}\nu^2(\mu) & = & 
\frac{\mu^{-2\epsilon}}{8\pi^2} (-3e^2+4\lambda+g_Y^2)\nu^2, \la{nurun} \\
\mu\frac{d}{d\mu}\lambda(\mu) & = & 
\frac{\mu^{-2\epsilon}}{8\pi^2} 
(3e^4-6\lambda e^2+10\lambda^2+2\lambda g_Y^2-g_Y^4),  \\
\mu\frac{d}{d\mu}g_Y^2(\mu) & = & 
\frac{\mu^{-2\epsilon}}{8\pi^2} 
(2g_Y^4-3e^2g_Y^2). \la{run4}
\ea

To relate the running parameters to the physical observables, 
one has to calculate suitable physical quantities to 1-loop order, 
and express the running parameters in terms of these. 
For illustration, let us consider relating the running parameter
$\nu^2(\mu)$ to physical masses. 
To do so, we calculate the physical pole mass $m_H$ 
to 1-loop order. For the calculation, the field $\phi_1$ is 
shifted to the classical broken minimum $\phi_c=\nu^2/\lambda$. 
Then the  radiatively corrected 
renormalized 1-loop propagator of the Higgs
field is of the form 
\be
\langle\phi_1(-k)\phi_1(k)\rangle=\frac{1}{k^2+m_1^2-\Pi_H(k^2,\mu)},
\ee
where $m_1^2=2\nu^2$.
Solving for the pole $m_H$, one gets
\be
\nu^2(\mu)=\frac{m_H^2}{2}\biggl[1+\frac{\Pi_H(-m_H^2,\mu)}{m_H^2}\biggr].
\label{nu2mu}
\ee
This expression is gauge-independent, since the 
self-energy is evaluated at the pole. The latter term in 
eq.~\nr{nu2mu} is the 1-loop correction to the tree-level
formula in eq.~\nr{0l}, and contains the $\mu$-dependence
in eq.~\nr{nurun}.

To give the 
explicit form of the 1-loop expression for $\nu^2(\mu)$,
we use the notation 
\be
h\equiv\frac{m_H}{m_W},\quad t\equiv \frac{m_t}{m_W}.
\ee
Inside 1-loop formulas, one may then write 
$\lambda=e^2h^2/2,\quad g_Y^2= 2 e^2t^2$.
Let us also introduce the function
\be
L_1(h)=-2\sqrt{4h^{-2}-1}\arctan\frac{1}{\sqrt{4h^{-2}-1}}, 
\ee
which has the special value $L_1(1)=-\pi/\sqrt{3}$.
Then we get 
\ba
{\nu}^2(\mu)&=& \frac{m_H^2}{2}\biggl\{
1+\frac{e^2}{16\pi^2}\biggl[  
(2h^2+2t^2-3)\ln\frac{\mu^2}{m_W^2} \nonumber \\
& & -3h^2\ln h + \Bigl(7-\frac{3\sqrt{3}}{2}\pi\Bigr)h^2
-5 + \frac{6}{h^2} 
-4 t^2\ln t+4 t^2-8 \frac{t^4}{h^2}
\nonumber \\
& & + \frac{h^4-4h^2+12}{2 h^2}L_1(h) 
+ 2 t^2 \Bigl(1-4\frac{t^2}{h^2}\Bigr)L_1(h/t)
\biggr]\biggr\}. \label{nu2ph}
\ea

\subsection{Integration over the superheavy scale}
\label{ioshs}

At high temperatures, the equilibrium thermodynamics 
of the theory defined by\linebreak eq.~\nr{4dl} can be described 
by a superrenormalizable 3d effective field theory
(for a review, see~\cite{erice}).
The Lagrangian of the effective theory is that in eq.~\nr{ac} with
a second scalar $A_0$:
\ba 
{\cal L}_{\rm eff} & = & 
\frac{1}{4}F_{ij}F_{ij}+
(D_i\phi)^*(D_i\phi)+
m_3^2\phi^*\phi+\lambda_3 (\phi^*\phi)^2 \nonumber \\
& & +\frac{1}{2} (\partial_iA_0)^2+\frac{1}{2}m_D^2A_0^2+
\frac{1}{4}\lambda_A A_0^4 +
h_3\phi^*\phi A_0^2,
\label{action}
\ea  
where all the fields have
the dimension GeV$^{1/2}$ and the couplings $e_3^2, \lambda_3, 
\lambda_A, h_3$ have the dimension GeV.

The purpose of this section is to give the parameters 
$e_3^2, h_3, m_3^2, \lambda_3, m_D^2, \lambda_A$ 
and the fields $A_i, \phi$ in terms
of the parameters in eq.~\nr{4dl}, with the 
accuracy compatible with 1-loop renormalization 
of the vacuum theory, as described above. 
Apart from terms arising from fermions, the results 
have been given at 1-loop order in~\cite{perturbative}
(2-loop results for the mass parameters have 
been added in~\cite{andersen2}). Here we add the 
effects of fermions. We use extensively the
generic results from~\cite{generic}. The basic notations are: 
\ba
& & c = \frac{1}{2}\biggl[\ln \frac{8\pi}{9}+\frac{\zeta'(2)}{\zeta(2)}-
2\gamma_E]\approx -0.348725, \nonumber \\ 
& & L_b(\mu) = 
2 \ln\frac{\mu e^{\gamma_E}}{4\pi T}\approx
2 \ln\frac{\mu}{7.0555T},\quad 
L_f(\mu)= 2 \ln\frac{\mu e^{\gamma_E}}{\pi T}\approx
2 \ln\frac{\mu}{1.7639T}. \mbox{\hspace*{1cm}}
\ea

Let us start with the relations of the wave functions.
Using the ${\cal Z}^{\phi}$'s from eqs.~(35--37) of~\cite{generic},
the momentum-dependent contribution of the superheavy modes
to the two-point scalar correlator is 
\be
{\cal Z}^{\phi}={\cal Z}^{\phi}_{\rm CT}-
e^2{\cal Z}^{\phi}_{\rm SV}+
\frac{1}{2}g_Y^2{\cal Z}^{\phi}_{\rm FF}.
\ee 
For the spatial and temporal components of the gauge fields one gets
\be
{\cal Z}^{A}={\cal Z}^{A}_{\rm CT}-e^2{N_s}{\cal Z}^{A}_{\rm SS}
-e^2N_f{\cal Z}^{A}_{\rm FF}, 
\ee
where the ${\cal Z}^{A}$'s are from eqs.~(38--45) of~\cite{generic}.
Hence the wave functions in the 3d action are related to 
the renormalized 4d wave functions in the $\overline{\rm MS}$ scheme by
\ba
\phi_3^2 & = &
\frac{1}{T} 
\phi^2(\mu) \biggl\{1+\frac{1}{16\pi^2}\biggl[-3e^2L_b(\mu)+
g_Y^2L_f(\mu)\biggr]\biggr\}, \label{phi3} \\
\bigl(A_0^{3d}\bigr)^2 & = &
\frac{1}{T}  
A_0^2(\mu)\biggl\{1+\frac{1}{16\pi^2}\frac{e^2}{3}\biggl[
N_s L_b(\mu)
+2 N_f L_f(\mu)+2 (N_s-N_f)\biggr]\biggr\}, \label{A03} \\
\bigl(A_i^{3d}\bigr)^2 & = &
\frac{1}{T}  
A_i^2(\mu)\biggl\{1+\frac{1}{16\pi^2}\frac{e^2}{3}\biggl[ 
N_s L_b(\mu)+2 N_f L_f(\mu)
\biggr]\biggr\}. \label{Ai3}
\ea

The couplings $e_3^2$ and $h_3$
can be extracted from the superheavy contributions to
the $(\phi\phi A_iA_j)$- and $(\phi\phi A_0A_0)$-correlators
at vanishing external momenta, respectively. 
The contributions from the relevant diagrams 
are in eqs.~(50--63) of~\cite{generic}.
The result is
\be
{\cal G}^{A}=
{\cal G}^{A}_{0}
+{\cal G}^{A}_{\rm CT}
-4\lambda e^2{\cal G}^{A}_{\rm SS}
-4e^4{\cal G}^{A}_{\rm SV} 
+4\lambda e^2{\cal G}^{A}_{\rm SSS}
-e^2g_Y^2{\cal G}^{A}_{\rm FFFF}.
\ee
When the fields are redefined according 
to eqs.~\nr{phi3}-\nr{Ai3} and the vertex is identified 
with the corresponding vertex in eq.~\nr{action}, one gets
\ba
h_3 & = & e^2(\mu)T
\biggl\{1+ \frac{1}{16\pi^2}\biggl[
-\frac{e^2N_s}{3} L_b(\mu)
-\frac{2e^2N_f}{3} L_f(\mu)+ \nonumber\\
&&\qquad\qquad\qquad +2 e^2 \Bigl(1+\frac{N_f-N_s}{3}\Bigr)
+8 \lambda-2 g_Y^2 \biggr]\biggr\}, \\
e_3^2 & = & e^2(\mu)T
\biggl\{1+\frac{1}{16\pi^2}\biggl[
-\frac{e^2N_s}{3} L_b(\mu)
-\frac{2e^2N_f}{3} L_f(\mu)
\biggr]\biggr\}. \label{g32}
\ea

For the scalar sector, the required correlators
are most conveniently generated from the effective potential.
With the Lagrangian masses 
\ba
&  & m_1^2 = -\nu^2+3\lambda\phi^2,\quad
m_2^2=-\nu^2+\lambda\phi^2,\nonumber \\
&  & m_T^2 = e^2\phi^2,\quad
m_f^2=\frac{1}{2}g_Y^2\phi^2, \label{4dms}
\ea
the 1-loop (unresummed) contribution to 
the effective potential in Landau gauge is 
\be
V_1(\phi)={\cal C}_S(m_1)+{\cal C}_S(m_2)+
{\cal C}_V(m_T)+ {\cal C}_F(m_f),\la{1looppot}
\ee
where the ${\cal C}$'s are from 
eqs.~(69--71) of~\cite{generic}. 
{}From the term quartic in masses in $V_1(\phi)$, 
one gets the superheavy contribution to the four-point scalar 
correlator at vanishing momenta. Redefining the 
field $\phi$ according to eq.~\nr{phi3},  
the coupling constant $\lambda_3$ is then  
\ba
\lambda_3 & = & T\biggl\{
\lambda(\mu)+
\frac{1}{16\pi^2}\biggl[\biggl(
-3e^4+6\lambda e^2-10\lambda^2\biggr)L_b(\mu)+\nonumber\\
&&\qquad\qquad +\biggl(g_Y^4-2\lambda g_Y^2\biggr)L_f(\mu)+
2 e^4 \biggr]\biggr\}. 
\ea
The coefficient of $\phi^2/2$ in $V_1(\phi)$ gives the 1-loop
result for the scalar mass squared. The result
is the term of order $g^2$ on the first line 
of eq.~\nr{m32}.

For the 2-loop contribution to the mass squared $m_3^2$, 
one needs the 2-loop effective potential $V_2(\phi)$.
In terms of eqs.~(81--93) of~\cite{generic}, the result is
\ba
V_2(\phi) & = & 
-\lambda^2\phi^2\Bigl[3{\cal D}_{\rm SSS}(m_1,m_1,m_1)
+{\cal D}_{\rm SSS}(m_1,m_2,m_2)\Bigr]
\nonumber \\ & &
-\frac{1}{2}e^2
{\cal D}_{\rm SSV}(m_1,m_2,m_T)
-\frac{1}{4}e^4\phi^2{\cal D}_{\rm SVV}(m_1,m_T,m_T) 
\nonumber \\ & &
-\frac{1}{4}g_Y^2\Bigl[
{\cal D}_{\rm FFS}(m_f,m_f,m_1)
+{\cal D}_{\rm FFS}(m_f,m_f,m_2)\Bigr]
-\frac{N_f}{2}e^2
{\cal D}_{\rm FFV}(m_f,m_f,m_T)
\nonumber \\ & &
-\frac{1}{4}\lambda\Bigl[
3{\cal D}_{\rm SS}(m_1,m_1)
+2{\cal D}_{\rm SS}(m_1,m_2)
+3{\cal D}_{\rm SS}(m_2,m_2)\Bigr]
\nonumber \\ & &
-\frac{1}{4}e^2\Bigl[
{\cal D}_{\rm SV}(m_1,m_T)
+{\cal D}_{\rm SV}(m_2,m_T)\Bigr]
\nonumber \\ & &
+\frac{1}{2}{\cal D}_{\rm S}(m_1)
+\frac{1}{2}{\cal D}_{\rm S}(m_2)
+\frac{1}{2}{\cal D}_{\rm V}(m_T)
+{\cal D}_{\rm F}(m_f).
\label{V2sum}
\ea
With the abbreviations
\ba
\tilde{\nu}^2 & = & \nu^2(\mu)\biggl\{1+\frac{1}{16\pi^2}
\biggl[
3e^2L_b(\mu)-4\lambda L_b(\mu)-g_Y^2L_f(\mu)
\biggr]\biggr\}, \label{tnu2} \\
\tilde{g}_Y^2 & = & T g_Y^2(\mu)\biggl\{1+
\frac{1}{16\pi^2}\biggl[
3e^2L_b(\mu)-2g_Y^2L_f(\mu)-e^2(1+2\ln 2)+\nonumber\\
&&\qquad\qquad\qquad+24 \lambda \ln 2-4 g_Y^2 \ln 2
\biggr]\biggr\} ,
\ea
the result for the 3d mass parameter is
\ba
m_3^2(\mu) & = & -\tilde{\nu}^2
+T\biggl(\frac{1}{3}\lambda_3+\frac{1}{4}e_3^2+
\frac{1}{12}\tilde{g}_Y^2\biggr)
\nonumber \\ 
& + &
\frac{T^2}{16\pi^2} \biggl[
\frac{1}{18}
e^4\biggl(N_f-16+18 N_f \ln 2 \biggr)+
\frac{2}{3}\lambda e^2 \biggr]
\nonumber \\
& + &
\frac{1}{16\pi^2} 
\biggl(-2h_3^2-4e_3^4+
8\lambda_3e_3^2-8\lambda_3^2\biggr)
\biggl(\ln\frac{3 T}{\mu}+c\biggr). 
\label{m32}
\ea
Here we have taken into account higher-order
corrections on the last line, by replacing the 4d coupling
constants by the 3d ones which appear in the exact running 
of $m_3^2(\mu)$ inside the superrenormalizable 3d theory.

The parameters $m_D^2$ and $\lambda_A$ require 
the calculation of the superheavy contributions on the 
$(A_0A_0)$- and $(A_0A_0A_0A_0)$-correlators 
at vanishing momenta.
Using eqs.~(96--101) of~\cite{generic}, the 1-loop 
2-point correlator for the U(1)-field is 
\ba
{\cal A}^{(2)} & = &
e^2\Bigl[2N_s{\cal A}^{(2)}_{\rm S}
-N_s{\cal A}^{(2)}_{\rm SS}
-N_f{\cal A}^{(2)}_{\rm FF}\Bigr]\label{a2u1}
\ea
and using eqs.~(102--109) of~\cite{generic}, the 
four-point correlator is
\ba
{\cal A}^{(4)}_{\rm SU(2)} & = &e^4\Bigl[
-2N_s{\cal A}^{(4)}_{\rm SS}
+4N_s{\cal A}^{(4)}_{\rm SSS}
-N_s{\cal A}^{(4)}_{\rm SSSS}
+N_f{\cal A}^{(4)}_{\rm FFFF}\Bigr].\label{a4su2}
\ea
Since there is no tree-level term corresponding to 
the correlators in eqs.~\nr{a2u1}-\nr{a4su2},
the redefinition of fields in eq.~\nr{A03} produces
terms of higher order. The final results
can then be read directly from eqs.~\nr{a2u1}-\nr{a4su2}:
\ba
m_D^2 & = & \frac{e^2 T^2}{6}(2N_s+N_f), \\ 
\lambda_A & = & \frac{e^4T}{6\pi^2} (N_s-N_f). 
\ea
Apart from fermionic contributions, 
the 2-loop result for $m_D^2$ has been given 
in~\cite{andersen2}.

Using eqs.~\nr{run1}-\nr{run4}, one 
sees that the quantities $e_3^2$, $h_3$,  
$\lambda_3$, $\tilde{\nu}^2$ and $\tilde{g}_Y^2$  
are independent of $\mu$ to the
order they are presented above. In other words,  
when the running parameters 
$e^2(\mu)$, $\nu^2(\mu)$, $\lambda(\mu)$
and $g_Y^2(\mu)$ are expressed in terms of physical
parameters, the $\mu$-dependence
cancels in the 3d parameters.
Note also that the bare mass parameter $m_3^2$ 
of the 3d theory
is independent of~$\mu$, so that one may use an independent 
renormalization scale inside the 3d theory.

\subsection{Integration over the heavy scale}
\label{iohs}
Consider now the parametric magnitude of the couplings in 
eq.~\nr{action}. If we are interested in the study of the phase
transition, we know this happens when $m_3^2(\mu)$ is very small and
we see from eq.~\nr{m32} that the tree-level term and the 1-loop term
have to cancel so as to leave $m_3^2\sim e^4T^2$. At the same time
the mass of $A_0$ is of the order of $eT$.
The action in eq.~\nr{action} can then be further simplified
by integrating out the $A_0$-field, leaving an action of the
G-L form in eq.~\nr{ac}. 

If we proceed from the $T\sim T_c$ region to $T\gg T_c$, 
eq.~\nr{m32} implies that also $m_3$ is of the order of
$eT$ and thus it should be integrated out together with $A_0$.
The resulting theory is very simple: free electrodynamics in 3d.
This is the statement that the magnetic sector of hot scalar ED
is trivial.

For the case $T\sim T_c$ when the action is the one in eq.~\nr{ac}, 
we denote the new parameters by a bar. The relations of
the old and the barred parameters have been given
in~\cite{perturbative}, and we include the results here just for 
completeness. 

The integration over the heavy scale proceeds
in complete analogy with integration over
the superheavy scale in Sec.~\ref{ioshs}. 
Since there are no momentum-dependent 1-loop 
corrections to the $\phi_3$- and $A_i$-fields
from the $A_0$-field, $\bar{\phi}_3=\phi_3$ and 
$\bar{A}_i = A_i^{3d}$. Since $A_0$ and $A_i$ do not interact, 
there are no 1-loop corrections to the 
$(\phi_3)^2 \bigl(A^{3d}_{i}\bigr)^2$-vertex, 
so that $\bar{e}_3=e_3$. The scalar self-couplings 
get modified, and at 1-loop order
\be
\bar{\lambda}_3=\lambda_3-\frac{1}{8\pi}\frac{h_3^2}{m_D}, \quad
\bar{m}_3^2= m_3^2-\frac{1}{4\pi}h_3 m_D.
\ee
To calculate the 2-loop corrections to the mass parameter,
we use again the effective potential. The 2-loop contribution from 
the heavy scale to the effective potential is
\ba
V_2^{\rm heavy}(\phi) = 
\frac{1}{2}h_3\Bigl[
D_{\rm LS}(m_1)+D_{\rm LS}(m_2)\Bigr]
-h_3^2\phi^2D_{\rm LLS}(m_1),
\ea
where the $D$'s are from eqs.~(119--122) of~\cite{generic}.
The mass parameter $\bar{m}_3^2(\mu)$ is then 
\ba
\bar{m}_3^2(\mu) 
& = & -\tilde{\nu}^2
+T\biggl(\frac{1}{3}\lambda_3+\frac{1}{4}e_3^2+
\frac{1}{12}\tilde{g}_Y^2\biggr)
-\frac{1}{4\pi}h_3 m_D \nonumber \\
& & \hspace*{0.5cm}
+\frac{T^2}{16\pi^2} \biggl[
\frac{1}{18}
e^4\biggl(N_f-16+18N_f \ln 2\biggr)+
\frac{2}{3}\lambda e^2\biggr]
\nonumber \\
& & \hspace*{0.5cm}
+\frac{1}{16\pi^2} 
\biggl[
\biggl(-2h_3^2
\biggr)\biggl(\ln\frac{3 T}{2 m_D}+c+\fr12\biggr) 
\biggr]
\nonumber \\
& & \hspace*{0.5cm}
+\frac{1}{16\pi^2} 
\biggl(-4\bar{e}_3^4+
8\bar{\lambda}_3\bar{e}_3^2-8\bar{\lambda}_3^2\biggr)
\biggl(\ln\frac{3 T}{\mu}+c\biggr), 
\label{bm32}
\ea
where we used eq.~\nr{m32}
and included higher order corrections on the last line, 
by replacing the couplings with those of the effective theory. 

Using the results of this section for $\bar m_3^2(\mu)$ and
$\bar\lambda_3$ and adding the running parameters from 
Sec.~\ref{ropms}, the final parameters $y$ and $x$
of the G-L action in eq.~\nr{ac} are completely
fixed in terms of the physical quantities $e^2$, the temperature $T$, 
and the physical pole masses $m_W, m_H$ and $m_t$ at zero
temperature. We do not write out the expressions explicitly.
For illustration, 
let us show the final expression for $\tilde{\nu}^2$
from eqs.~\nr{nu2ph}, \nr{tnu2}:
\ba
\tilde{\nu}^2 &=& \frac{m_H^2}{2}\biggl\{
1+\frac{e^2}{16\pi^2}\biggl[ 
2(3-2h^2)\ln\frac{m_We^{\gamma_E}}{4\pi T}-
4 t^2\ln\frac{m_We^{\gamma_E}}{\pi T} \nonumber \\
& & -3h^2\ln h + \Bigl(7-\frac{3\sqrt{3}}{2}\pi\Bigr)h^2
-5 + \frac{6}{h^2} 
-4 t^2\ln t+4 t^2-8 \frac{t^4}{h^2}
\nonumber \\
& & + \frac{h^4-4h^2+12}{2 h^2}L_1(h) 
+ 2 t^2 \Bigl(1-4\frac{t^2}{h^2}\Bigr)L_1(h/t)
\biggr]\biggr\}.
\ea

\end{document}